\documentclass[12pt,a4paper,twoside,final,notitlepage, leqno]{article}
\usepackage[english]{babel}
\usepackage[T1]{fontenc}  
\usepackage{graphicx}
\def\br {\break}

\usepackage{amsmath,amsthm,epsfig,amsfonts,amssymb,bbm}  

\newcommand{\pequationdeb}{$$ \left\{ \begin{minipage}[c]{130mm}}
\newcommand{\pequationfin}{\end{minipage}
                           \right. $$}

\def \smb {{\scriptstyle \bullet }}
\newcommand{\monitem}{ \smallskip \noindent $\bullet$ \quad  } 
\newcommand{\moneq}{\vspace*{-6pt} \begin{equation} \displaystyle } 
\newcommand{\moneqstar}{\vspace*{-6pt} \begin{equation*} \displaystyle } 
\newcommand{\monendstar}{\vspace*{-2pt} \end{equation*}   }
\newcommand{\monend}{\vspace*{-6pt} \end{equation}   }

\def\R{{\rm I}\! {\rm R}}

\def\section*#1{}

\def\resume{\if@twocolumn
\section*{R\'esum\'e}
\else \small
\quotation{\bf \it R\'esum\'e \rule[1mm]{1.5mm}{0.2mm}\vspace{0pt}}
\fi}
\def\endresume{\if@twocolumn\else\endquotation\fi}
\def\abstract{\if@twocolumn
\noindent\section*{{\bf Abstract}}
\else \small
\quotation{\noindent \bf {Abstract.} \rule[1mm]{1.5mm}{0.2mm}\vspace{0pt}}
\fi}
\def\endabstract{\if@twocolumn\else\endquotation\fi}

\usepackage{fancyhdr}
\begin{document}

~

\bigskip \bigskip   \bigskip

\centerline {\bf \LARGE   Coupling Linear Sloshing with Six Degrees  }

\bigskip   \centerline {\bf \LARGE   of Freedom  Rigid Body Dynamics }

\bigskip \bigskip \bigskip \bigskip

\centerline 
{ \large  Fran\c{c}ois Dubois$\,^{ab*}$,  Dimitri Stoliaroff$\,^{c}$  and Isabelle Terrasse$\,^{d}$}   

\bigskip  
\centerline { \it  \small    $^a$   Conservatoire National des Arts et M\'etiers,  Paris, France.  } 
\centerline { \it  \small    $^b$     Department of Mathematics, University Paris Sud, Orsay,  France.  } 
\centerline { \it  \small    $^c$     Airbus Defence and Space,   Les Mureaux,  France.  } 
\centerline { \it  \small    $^d$     Airbus Group Innovations, Suresnes, France.  } 
\centerline { \it  \small    $^*$     corresponding author  }

\centerline { \it  \small  francois.dubois@cnam.fr,  dimitri.stoliaroff@astrium.eads.net, 
 isabelle.terrasse@airbus.com } 
   \bigskip   

\bigskip \bigskip
 
\centerline { 13 july 2015~\footnote{~Contribution published  
 in {\it European Journal of Mechanics-B - Fluids}, 
doi:10.1016/j.euromechflu. 2015.06.002, volume~54, pages~17-26, november-december 2015.  }}

\bigskip

\noindent {\bf Abstract }  

Fluid motion in tanks is usually described in space industry 
with the so-called Lomen hypothesis which assumes the vorticity
is null in the moving frame. We establish in this contribution that this hypothesis 
is valid only for uniform rotational motions. 
We give a more general
formulation of this coupling problem, with a compact formulation.

We consider the   mechanical modeling of a 
 rigid body with a  motion of small amplitude,  containing an
 incompressible fluid in the linearized regime. 
We first establish that  
the fluid motion remains irrotational in  a Galilean referential
if it is true at the initial time.   
When  continuity of normal velocity and  pressure are prescribed on the 
free surface, we establish that the global coupled problem 
conserves an energy functional composed by three terms.  
We introduce 
 the Stokes - Zhukovsky  vector fields,  
solving   Neumann problems for the Laplace operator in the fluid
in order to represent the rotational rigid motion with irrotational vector fields.
Then   we have a good   framework to consider the coupled problem   
between the fluid and the rigid motion. 
The coupling between the free surface and the {\it ad hoc}  component of the velocity
potential introduces a ``Neumann to Dirichlet'' operator that allows to write the coupled
system in a very compact form.
The final expression of a Lagrangian for the coupled system is derived 
and the Euler-Lagrange equations of the coupled motion are presented.  

\smallskip \noindent 
   {\bf Keywords}: 

\hfill Stokes,  Zhukovsky, Fraeijs de Veubeke,    vector fields, 
 integral boundary operator.


\newpage

\fancyfoot[C]{\oldstylenums{\thepage}}   
\fancyhead[EC]{\sc{Fran\c{c}ois Dubois, Dimitri Stoliaroff and Isabelle Terrasse}} 
\fancyhead[OC]{\sc{Coupling Linear Sloshing with Rigid Body Dynamics}}  

\bigskip \bigskip   \noindent {\bf \large   Scope of the problem}   

\noindent
Sloshing of liquid in tanks is an important phenomenon 
for space and  terrestrial applications. 
We think for example of   sloshing effects in 
road vehicles and   ships carrying liquid cargo. 
The question is to know the magnitude  of the wave
and the total effort on the structure due to the movement of the fluid. 
For this kind of problematics, a lot of references exist and we 
refer the reader {\it i.e.} to the book of H.~Morand and R.~Ohayon 
\cite{MO92},  to the review proposed by
R.~Ibrahim, V.~Pilipchuk and  T.~Ikeda \cite {IPI01}, 
 to the book of R.A.~Ibrahim \cite {IPI05},   
the book of O.M.~Faltinsen and A.N.~Timokha \cite{FRLT09}
or to the review article of  G.~Hou  {\it et al.}  \cite{HWL12}.

%
\smallskip \noindent
Moreover, for industrial applications, we would have a movable rigid tank with 
liquid free surface  with six possible rigid  movements  
and without needing a complete study of the  elastic body, 
as studied {\it e.g.} in  H.~Bauer {\it et al.}   \cite{BHW68}, 
S.~Piperno  {\it et al.}  \cite{PFL95}, 
C.~Farhat  {\it et al.}  \cite{FLT98}, 
 J.F.~Gerbeau and  M.~Vidrascu \cite{GV03}, 
K.J.~Bathe and H.~Zhang \cite{BZ04}, 
T.E.~Tezduyar  {\it et al.}  \cite{TSSA06} 
and  the previous references. 

\smallskip \noindent
In our case relative to space applications, the fundamental hypothesis
of this contribution 
is the existence of some propulsion. We do not consider in this study 
the very complicated and nonlinear movement due to the quasi-disparition
of gravity field. 
We refer for such  studies to the contributions 
of  F.~Dodge and L.~Garza \cite{Do70, DG67},  
S.~Ostrach  \cite{Os82},   H.~Snydera \cite{Sn82}, 
C.~Falc\'on {\it et al.}  \cite{FFBF09} 
and 
P. Behruzi, {\it et al.}   \cite {BRNS11} 
among others.  
%
On the contrary, a gravity field is supposed to be present 
in our contribution and 
moreover an extra-gravity field is added due to the propulsion system. 
Then it is legitimus to linearize all the geometrical deformations and 
the equations of dynamics. 
In this kind of situation, the knowledge of the action of the fluid on the
structure is mandatory. The question has been intensively studied during the sixties
under the impulsion of NASA 
(see {\it e.g.} 
H.~ Bauer  \cite {Ba64}, 
D.~Lomen \cite {Lo64a, Lo64b},  
H.~Abramson  \cite {NA66}, 
L.~Fontenot   \cite {Fo68}) 
 and in European countries in the seventies 
(see  {\it e.g.}   J.P.~Leriche  \cite {Le72})
or in the context of Ariane~5 studies (B.~Chemoul {\it et al.} \cite {CLRSP01}). 

\smallskip \noindent
We observe that due to its own intrinsic movement, 
the structure has also some influence on the fluid displacement.
This question has been rigorously studied by the Russian school
in the sixties (N.~Moiseev and  V.~Rumiantsev \cite{MR65}).
It is sufficient in a first approach to consider the solid as a rigid body and 
to neglect all the flexible deformations. 

\smallskip \noindent
In fact, we are in front of a  complete coupled problem. 
The fluid is linearized and has an action on the 
solid, considered as a rigid body. 
The solid is a   ``six degrees of freedom'' system 
that can also be considered as linearized around 
a given configuration. 
This coupled problem does not seem to have been considered
previously under this form in the literature. 
We observe that this quite old problem
raises  actually an intensive scientific activity.
As examples, we mention the   contributions of 
O.~Faltinsen, O.~Rognebakke,  I.~Lukovsky and A.~Timokha    \cite{FRLT2k, GHTY10}  
who derived a variational method to analyze the sloshing with finite water depth.  
Note also that 
K.~London  \cite {Lon01}  
analyzed the case of a multi-body model with applications to
 the Triana spacecraft, and 
%
J.~Vierendeels {\it at al.}   \cite {VDDV05} 
proposed to use the Flow3D computer software (Fluent, Inc)  to analyze 
numerically nonlinear effects involved in the coupling of a rigid body with sloshing fluid, 
L.~Diebold {\it at al.}    \cite{DBHZ08} 
studied  the effects on sloshing pressure due to the coupling between 
seakeeping and tank liquid motion.  
In the  thesis of  A.~Ardakani,  
the general rigid-body motion with interior
shallow-water sloshing is studied in great detail and  
we refer to the communication of  A.~Ardakani 
and   T.~Bridges   \cite {AB10}. 
A time-independent finite difference method
to solve the problem of sloshing waves and resonance modes of fluid 
in a tridimensional tank is also considered by 
C.~Wu and  B.~Chen  \cite{Ch09}.

\smallskip \noindent 
We begin this article with classical  considerations on sloshing in a fixed solid.
 We focus  on the free surface 
and to  usual  physical ingredients: the continuity of normal velocity and 
the continuity of pressure. 
The coupling between the free surface and 
the velocity potential introduces a ``Neumann to Dirichlet'' integral operator 
that allows to write  the coupled system in a very compact form.
Then in Section~2, we recall 
fundamental aspects of the dynamics of a six  degrees  of freedom  rigid body dynamics:
description of the rigid body and its infinitesimal motion, 
the incompressible fluid and its linearization. We discuss  
the so-called ``Lomen hypothesis'' intensively used for industrial space applications
and prove that,  with a good generality, the fluid motion remains irrotational in 
a Galilean referential. 
We introduce some special vectorial functions
that we call the ``Stokes - Zhukovsky vector fields'', 
independently rediscovered by  
multiple generations of great scientists during the two last centuries
(see {\it e.g.} G.~Stokes \cite{St1843},  N.~Zhukovsky \cite{Zh1885} and  
 B.~Fraeijs de Veubeke \cite{dV63}). 
These vector fields solve Neumann problems for the Laplace operator in the fluid  
and allow the representation of a rigid body displacement by an irrotational field. 
It is a good framework to consider the coupled problem.  
Then the dynamics equations of the rigid body in the presence of an internal sloshing fluid 
are established. 
In Section~3, we study the coupled problem. 
We do not incorporate any dissipation
and in consequence we establish the conservation of energy for
this simple case. We propose a compact set of variables to describe the entire 
coupled dynamics. 
Then the  coupled system appears 
in a very simple form formally analogous to a scalar harmonic oscillator! 
Finally, the  expression of a Lagrangian for the coupled system is proposed.

\smallskip \noindent 
When the explanations of the mathematical results are not detailed, 
we refer the reader to the classical books 
of N.~Moiseev and  V.~Rumiantsev \cite{MR65}, 
H.~Morand and R.~Ohayon  \cite{MO92},  
R.A.~Ibrahim \cite {IPI05},  
O.M.~Faltinsen and A.N.~Timokha \cite{FRLT09}
or to the preliminary edition 
\cite{DS14} of this contribution. 

\bigskip \bigskip   \noindent {\bf \large 1) \quad  Sloshing in a fixed solid  }  

 \noindent  
In this section, the studied mechanical system  is the fluid.
The liquid is contained inside the  solid $\cal S$, it  
occupies a volume $\, \Omega(t) \,$   variable with   time, 
  with a    {\bf constant} density $\rho_L$. 
The total   mass   $ \, m_L \,$ of  liquid 
is the integral of the density   $ \, \rho_L \, $ on the volume  $\, \Omega(t) . \,$
%
At the boundary $\partial  \Omega $ of  liquid, we have a   
contact surface $\, \Sigma (t) \,$ between  liquid and   solid 
and $ \,  \Sigma (t) =  \partial  \Omega  \, \cap \,  \partial S $, as described in 
Figure~1,    
%
%
and a free surface  $\, \Gamma(t) \,$ where the  liquid is in thermodynamical 
 equilibrium with its vapor. 
%
%
The liquid is submitted to a  gravity field $ \, g_0 . \, $ This vector is collinear
to an ``absolute'' vertical direction associated with  a vector $ \, e_3 , \,$ 
third coordinate of a Galilean referential  $ \, (e_1 ,\, e_2 ,\, e_3) $: 
\moneqstar 
  g_0 \,=\, - g \, e_3  \, . 
\monendstar    
Note that $ \, g > 0 \, $ with this choice, as illustrated in Figure 1.  
The velocity field of the  liquid $\, u(t)\,$ 
is   measured relatively to an  absolute referential, following 
{\it e.g.} the work of L.~Fontenot \cite{Fo68}.
The  liquid is assumed incompressible:  
\moneq  \label{vitesse-liquide} 
 {\rm div} \, u  \,=\, 0 
\,\quad {\rm in} \quad \Omega  (t) \, . 
\monend
%

\bigskip 
\centerline { \includegraphics[width=.50 \textwidth]  {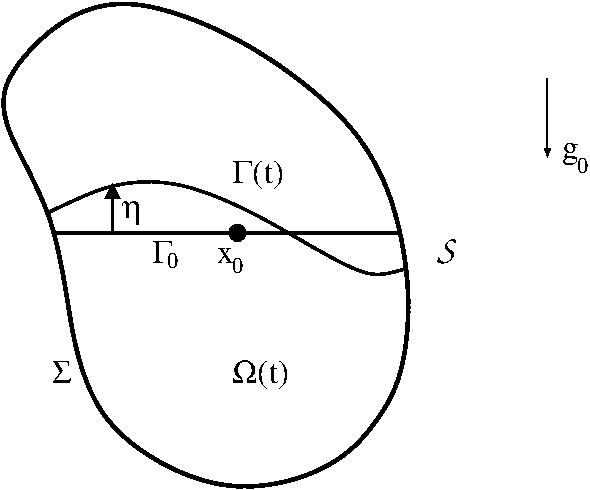} }  

\smallskip \noindent  {\bf Figure 1}. \quad 
General view of the sloshing problem in a solid at rest. 
The free boundary $ \, \Gamma (t) \, $ 
is issued from the 
equilibrium free boundary at rest  $ \, \Gamma_0 \, $
 with the help of the elongation~$ \, \eta . \, $

\smallskip   \monitem   {\bf      Liquid as a perfect linearized fluid  } 
  

\noindent 
The pressure field $ \, p(x) \, $ is defined in the liquid domain 
$ \,   \Omega \ni x \, \longmapsto \, p(x) \in \R . $ 
The conservation of momentum  for a  perfect fluid
is written with the  Euler equations of hydrodynamics:
%
$ \,  {{\partial u}\over{\partial t}} \,+\, ({\rm curl} \,u )  
\,  \times \, u  \,+\, 
\nabla \big( {{p}\over{\rho_L}} + {1\over2} \, \mid\! u \! \mid^2 \!\big)  \,=\, 
g_0 \, $ in $ \, \Omega(t)  . $ 
%
In this contribution, we 
 make a {\bf linearization hypothesis}. In particular, we neglect the nonlinear terms
in fluid dynamics  and 
 replace  the  previous equation by:
\moneq  \label{Euler-linear}    
 {{\partial u}\over{\partial t}} \,+\,  {{1}\over{\rho_L}} \nabla p 
 \,=\,  g_0 \, \quad {\rm in} \quad \Omega(t)  \,.
\monend

\smallskip   \monitem   {    \bf      Velocity potential  } 
 
  \noindent  
We suppose moreover that the fluid is irrotational:
\moneq   \label{rot-nul}   
  {\rm curl} \, u \,=\, 0  \,.
\monend 
If  the  domain $ \, \Omega(t) \, $ is simply connected 
 (be {\bf careful} with this hypothesis for  {\bf toric}  geometries !), 
the simple hypothesis (\ref{rot-nul}) implies that 
the velocity field can be generated by a potential~$ \, \varphi $:
\moneq   \label{potentiel}    
 u(x) \,= \, \nabla \varphi (x)  \,, \quad x \in \Omega(t) \, 
\monend 

\monitem
In order to have precise information  concerning this velocity potential, 
we recall the   Bernoulli theorem.  
We inject the  velocity field $ \,  u  \,= \, \nabla \varphi \, $ 
in the dynamical  equations. 
We introduce a point $\, P \,$: 
$ \,  \nabla \, \big( 
{{\partial \varphi}\over{\partial t}} \,+\, {{p}\over{\rho_L}} \,-\, 
g_0 \, \smb \, (x - x_P)  \big)   \,=\, 0  \,\,$ for $ \, x  \in \Omega  .$
%
We add some time function to the  scalar potential  
of  velocity   (and assume that the  domain $ \, \Omega \,$ is connected). 
Then: 
\moneq    \label{bernoulli}       
{{\partial \varphi}\over{\partial t}} \,+\, {{p}\over{\rho_L}} \,-\, 
g_0 \, \smb \, (x - x_P)     \,=\, 0  \,, \qquad x \in \Omega . \,
\monend 

\monitem
We take now into consideration the incompressibility hypothesis 
(\ref{vitesse-liquide}) together with the potential representation of
the velocity field (\ref{potentiel}). We then obtain  the 
Laplace equation:
\moneq    \label{laplace} 
\Delta  \varphi  \,=\, 0 \,\quad  {\rm in} \,  \Omega  (t) .\, 
\monend 
A first  boundary condition for this equation is a consequence of 
the continuity of the normal velocity $ \, u \smb n \,$ at the  
interface $\, \Sigma \,$ between solid and liquid:
\moneq    \label{neumann-solide-zero}  
    {{\partial \varphi}\over{\partial n}} \,= \, 0 \,, \qquad x \in \Sigma (t) \, . \,
\monend 
%

\smallskip    \monitem   {    \bf    Free surface } 
 
  \noindent  
Consider as a reference situation the  solid at rest.  
Then the  free surface at equilibrium has a given position $ \, \Gamma_0 \,$ 
as presented in Figure~1. 
We note  $ \,\,  \eta \, n_0 \,\,$ 
the   displacement of the  free boundary  
at position $ \, y  \in  \Gamma_0 ,\,$ 
 where  $\,  n_0 \,$ denotes the  outward normal direction to $\, \Gamma _0 $. 
In this case of a fixed solid, we have $ \, n_0 = e_3 $. 
The  point $x$   new position  
 takes  into account the   variation of the  free surface: 
\moneq    \label{point-surface-libre}   
  x \,=\, y  \,+\, \eta(y) \, n_0 \,, \qquad y \in \Gamma_0 \,, \quad x \in \Gamma .\,
\monend
We denote by $ \, x_0 \, $  the center of gravity of the  frozen free surface $\, \Gamma_0\,$: 
\moneq  \label{centre-surface-libre}   
\int_{ \Gamma_0 } \big( y - x_0 \big) \, {\rm d}\gamma \,=\, 0 \, . \, 
\monend 
Note that due to incompressibility condition, we have:
\moneq    \label{integrale-nulle}   
 \int_{\Gamma_0} \eta \,\, {\rm d}\gamma \,\equiv\, 0 \, .  
\monend  
Moreover, with $\, x \,$  given on $\, \Gamma(t) \,$ according to 
(\ref{point-surface-libre}), we have, 
\moneqstar    
\int_{ \Gamma_0 } ( x - x_0 ) \, {\rm d}\gamma \,=\, 0 \, . 
\monendstar 
We observe that, thanks to (\ref{integrale-nulle}),  
$ \, \int_{ \Gamma_0 } ( x - x_0 ) \, {\rm d}\gamma \, $ 
$\, = \, \int_{ \Gamma_0 } ( y - x_0   \,+\, \eta(y) \, n_0 ) \, {\rm d}\gamma $\br  
$\, = \, \int_{ \Gamma_0 } ( y - x_0 ) \, {\rm d}\gamma \,$ 
$\, = \, 0 \,$ due to the definition  (\ref{centre-surface-libre}). 
We introduce also the coordinates $\, X_1, $ $\, X_2, $ $\, X_3, $ of a point $x$
in the referential $ \, (x_0, \, e_1, e_2, e_3) $: 
$ \, x - x_0  \,=\, X_1 \, e_1 \,+\,  X_2 \, e_2 \,+ \,  X_3 \, e_3 $.

\bigskip   \noindent $\bullet$ \quad    {    \bf Proposition 1.  Neumann boundary
 condition   on the free surface     } 

\noindent
If we keep only the first order linear terms, the boundary condition for the velocity potential
on the free surface can be written as a kinematic condition:
\moneq  \label{neumann-surface-libre}   
 {{\partial \varphi}\over{\partial n }} \,=\, 
 {{\partial \eta}\over{\partial t }}  \,,  \qquad x \in  \Gamma_0 \, .\,
\monend 
Then  equations   (\ref{neumann-solide-zero}) and 
(\ref{neumann-surface-libre}) can be written in a synthetic form:
\moneq \label{potentiel-neumann}  
 {{\partial \varphi}\over{\partial n }} \,=\,  
   \left\{ \begin{array}{rcl}      \displaystyle 
    0  \,\,   &{\rm on }& \,\,  \Sigma   \\ 
 \displaystyle  
 \,\,   {{\partial \eta}\over{\partial t}}  \,  &{\rm on }& \,\, \Gamma_0   \, . 
\end{array}  \right.    \monend

\bigskip   \newpage \noindent $\bullet$ \quad    {    \bf  Proof of Proposition 1. } 

\noindent We introduce the equation $ \, F(X_1, \, X_2, \, X_3, \, t) = 0 \,$ 
of the free surface. We take the total derivative relative to time of this  constraint 
and replace the velocity $ \, {{{\rm d}X}\over{{\rm d}t}} \, $ by
the gradient $ \,\nabla \varphi \,$ of the potential. We obtain  
$ \,\, \nabla F \, \smb \, \nabla \varphi  +  {{\partial F}\over{\partial t}} \, = \, 0  $. 
%
The normal vector $ \, n \, $ can be written as 
$ \, n \, = \, \nabla F \, / \,\mid \! \nabla F \! \mid \, $
%
and we have 
$ \,   {{\partial \varphi}\over{\partial n}}  \equiv \nabla \varphi  \, \smb \, n \, $ 
$   \, = \, \nabla \varphi  \, \smb  \, {{\nabla F}\over{ \mid \! \nabla F \! \mid}}  \, . \, $
The previous equation 
can be written as
\moneq \label{prop1-1}  
{{\partial \varphi}\over{\partial n}} \, +  \, 
{{1}\over{\mid \! \nabla F \! \mid}} \, {{\partial F}\over{\partial t}} \, = \, 0  \, . 
\monend
We parameterize the surface  with an explicit function $ \, \eta ,\,$ {\it id est} 
\moneq \label{prop1-2}  
  F(X_1, \, X_2, \, X_3, \, t) \, \equiv \, X_3 -  \eta (X_1 ,\, X_2, \, t) .\,
\monend
Then linearizing the problem, we suppose that the free surface is close to 
its reference value $ \, \Gamma_0 \,$ at rest and 
we can neglect the gradient $ \, \nabla \eta \, $ of the free
surface equation (\ref{prop1-2}) compared to the unity. Thus we have 
$ \,  \mid \! \nabla F \! \mid \,= \, 1  \,+\, {\rm O} (\vert \eta\vert^2) $.  
%
Due to the particular form (\ref{prop1-2}), we deduce that we have 
$ \,   {{\partial F}\over{\partial t}} \,=\, -  {{\partial \eta}\over{\partial t}} \,$
on the free boundary 
and the relation (\ref{neumann-surface-libre}) is a direct consequence 
of (\ref{prop1-1}) and the fact that the norm of $\, \nabla F \,$ 
is of order unity. 


\bigskip   \noindent $\bullet$ \quad    {    \bf Proposition 2.  
Pressure continuity      across the free surface     } 

\noindent 
On the free surface $ \,  \Gamma ,\,$ the    
continuity of the  stress tensor   can be written for a perfect fluid
as a dynamic condition:
\moneq  \label{pression-surface-libre-zero}  
p \,=\, 0  \qquad {\rm on} \,\,  \Gamma \, . \, 
\monend 
It takes the following linearized form: 
\moneq  \label{pression-surface-libre-ini}    
 {{\partial \varphi}\over{\partial t }} \,+\,
g \, \eta 
 \, = \,  0 \,, \qquad x \in \Gamma_0   \, .
\monend 
%

\bigskip \noindent $\bullet$ \quad    {    \bf  Proof of Proposition 2. } 

\noindent 
The proof is classical and is explained  in classic  books 
as \cite{FRLT09, IPI05, MO92}. We give it here for completeness of the study. 
We choose  the   point $\, P  \,$ for the  Bernoulli equation  (\ref{bernoulli})
 on the  frozen free surface $\, \Gamma_0\,$  equal to the center 
$\, x_0 \, $ introduced in  (\ref{centre-surface-libre}).  
%
Due to Bernoulli theorem  (\ref{bernoulli}) and continuity 
of the pressure on  $\, \Gamma $, we deduce the following relation 
on the free surface: 
\moneq  \label{pression-surface-libre-initial}    
 {{\partial \varphi}\over{\partial t }} \,-\, g_0 \,\smb \, (x - x_0)  \,=\, 0 
 \,, \quad x \in \Gamma(t) .\,
\monend
%
We have the following calculus:
$ \,\, -\, g_0 \,\smb \, (x - x_0)  \,= $ 
$ \, g\,  {\rm e}_3  \,\smb \, \big(   X_1 \,  e_1  \,+\,  X_2  \,  e_2 
\,+\, \eta \, \,  {\rm e}_3 \big) \, = $ 
$ \,   g \,  \eta   $, 
%
and the condition   $ \, p = 0 \,$  
of pressure continuity  on  the free surface is expressed by
 $ \,  {{\partial \varphi}\over{\partial t }} \,+\,
g \,  \eta \, = \,  0  \, $  
which is exactly relation (\ref{pression-surface-libre-ini}).
The proof is established. 
\hfill  $ \square $ 

\smallskip    \monitem   {\bf     Free surface potential and  Neumann to Dirichlet operator }  
 
\noindent 
We introduce the ``free surface potential'' 
$ \, \Omega \ni x \,\longmapsto \, \psi(x) \in \R \,  $ satisfying  
the following Neumann boundary-value problem for the Laplace equation:
\moneq \label{pb-psi}      \left\{ \begin{array}{rcl}
\displaystyle \Delta   \psi &\,=\,&  0    \qquad {\rm in} \,\,   \Omega   \\ 
 \displaystyle     {{\partial \psi}\over{\partial n}}  &\,=\, & 
 \left\{ \begin{array}{rcl}
0   & {\rm on} &   \Sigma   \\
 \eta   & {\rm on} &     \Gamma_0   \, . 
\end{array}   \right. \end{array}    \right.    \monend 
\noindent 
We consider a free surface $ \, \eta \,$ such that the global incompressibility condition 
(\ref{integrale-nulle}) holds. 
We introduce the functional space  
\moneqstar 
F^{1/2} (\Gamma_0) \,\equiv \, \bigg\{ \, \eta : \Gamma_0 \longrightarrow \R , \,\,
\int _{\Gamma_0} \eta \, {\rm d}\gamma = 0 \, \bigg\} \, . 
   \monendstar 
We consider the ``free surface potential'' $ \, \psi \,$ associated
 to a given   $\, \eta \in F^{1/2} (\Gamma_0) \, $  in   the following way. 
The  function $ \,\,  \Omega \ni x \,\longmapsto \, \psi(x) \in \R \,\,   $ 
is uniquely defined by the Neumann problem (\ref{pb-psi}) with the additional condition    
\moneq \label{jauge-psi}    
\int _{\Gamma_0} \psi \, {\rm d}\gamma \, = \, 0  \, .
\monend 
We consider the restriction $ \, \zeta \,$ (the trace) 
of the function $ \psi$ on the surface $\Gamma_0$  
\moneq \label{restriction}    
  \, \Gamma_0 \ni x \,\longmapsto \, \zeta(x) \, \equiv \, \psi(x) \in \R \,  
\monend 
The mapping     $ \,\, 
 F^{1/2} (\Gamma_0) \ni \eta \longmapsto \zeta \in  F^{1/2} (\Gamma_0) \,\, $ 
 is the ``Neumann to Dirichlet'' 
operator. We denote it with the letter W: 
\moneq \label{def-W}    
 \zeta \, \equiv \, W  \smb\,  \eta \, . 
\monend 
A precise mathematical definition of the space $ \,  F^{1/2} (\Gamma_0) \,$ 
in the context of Sobolev spaces can be found  in \cite{LM68} or \cite{Ne01}. 
  
%
\bigskip   \noindent $\bullet$ \quad    {    \bf Proposition 3.  Positive  
self adjoint operator } 

\noindent
The operator $W$  $  \, : \, 
F^{1/2} (\Gamma_0) \ni \eta \, \longmapsto \, \zeta  \in F^{1/2} (\Gamma_0)  \,\, $
with  $ \zeta $ defined by the relations  (\ref{pb-psi}), (\ref{jauge-psi}), 
(\ref{restriction}) and (\ref{def-W}) is  self-adjoint. If we denote by 
$ \, (\smb ,\, \smb) \,$  the ${\rm L}^2$ scalar product on the linearized free surface 
$ \, \Gamma_0 ,\,$ {\it id est} 
\moneq \label{produit-scalaire} 
( \eta ,\, \zeta) \,\equiv\, \int_{\Gamma_0} \eta \,\, \zeta \, {\rm d}\gamma \,,  
\qquad \eta, \, \zeta \in   F^{1/2} (\Gamma_0) \, ,   
\monend 
we have: 
$ \,\,  ( \eta' \,,  W \smb\,  \eta ) \,=\,  (  W \smb\, \eta' \,,    \eta ) $, 
for all  $ \, \eta ,\, \eta' \in F^{1/2} (\Gamma_0) $.   
In particular, with the free surface potential $ \, \psi \,$ defined in 
(\ref{pb-psi}), we have 
\moneq \label{W-is-positive} 
 ( \eta \,,  W \smb\,  \eta ) \,=\, 
\int_{\Omega}  \big( \nabla  \psi \,\smb \,  \nabla \psi )  \,  {\rm d}x  \, \geq \, 0 \,  . 
\monend

%
\bigskip   \noindent $\bullet$ \quad    {    \bf  Proof of Proposition 3. }

\noindent 
We have with the previous notations: 

\smallskip \noindent 
$ \displaystyle \, ( \eta' \,,  W \smb\,  \eta ) \,=\, 
\int_{\Gamma_0} \eta' \, \psi \,  {\rm d}\gamma  \,=\,  
\int_{\partial\Omega} {{\partial \psi'}\over{\partial n}}  \, \psi \,  {\rm d}\gamma $
\hfill  because $  \displaystyle \, {{\partial \psi'}\over{\partial n}}  
= 0 \, $  on $\, \Sigma \, $  and  
$  \displaystyle \, {{\partial \psi'}\over{\partial n}}  
= \eta' \,$  on $\, \Gamma_0 \, $ 

\smallskip \qquad \quad  $ \displaystyle \, = \, 
\int_{\Omega} {\rm div} \big( \psi \, \nabla \psi' )  \,  {\rm d}x   $ 
$\hfill$ due to Green formula 

\smallskip \qquad \quad  $ \displaystyle \, = \, 
\int_{\Omega}  \big( \nabla  \psi' \,\smb \,  \nabla \psi )  \,  {\rm d}x     $
$\hfill$ because $ \,\, \Delta \psi' = 0 $ 

\smallskip \qquad \quad $ \displaystyle \, = \,   
\int_{\partial\Omega} \psi' \,  {{\partial \psi}\over{\partial n}}  \,   {\rm d}\gamma    $
$\hfill$ because $ \,\, \Delta \psi = 0 $ 

\smallskip \qquad  \quad $ \displaystyle \, = \,    
\int_{\Gamma_0}   \psi' \, \eta \,  {\rm d}\gamma        $
$ \displaystyle \, = \, 
 (  W \smb\, \eta' \,, \,    \eta ) \, . $ 

\smallskip \noindent We observe that if $ \, \eta = \eta' $, then
$ \, \psi = \psi' \, $ and the scalar product 
$ \,  (  W \smb\, \eta \,, \,    \eta ) \, $ is given by 
the relation (\ref{W-is-positive}) and is positive. 
The proof is complete. 
$\hfill  \square $ 

\monitem
In conclusion of the section, the velocity potential 
$\, \varphi \,$ satisfies (\ref{potentiel-neumann}), then its value on 
the free boundary $ \, \Gamma \,$ may be described in terms of the operator $W$. We have 
\moneqstar 
\varphi \,= \, W \, \smb\,  {{\partial \eta}\over{\partial t}} \,
\qquad  {\rm on} \,\, \Gamma_0 \, . 
\monendstar 
Then the evolution equation (\ref{pression-surface-libre-ini}) can be formulated 
only in terms of the {\it a priori} unknown free surface, parameterized by the function $\eta$ 
and the 
operator $W$: 
\moneq \label{sloshing-classique} 
\rho_L \,  W \, \smb\,  {{\partial^2 \eta}\over{\partial t^2}} \,+ \, \rho_L \, g \, \eta \,=\, 0 
\qquad  {\rm on} \,\, \Gamma_0 \, . 
\monend
This equation is usually presented  as a family of harmonic oscillators, 
through a diagonalization of the 
operator $ \, W \, $ with usual spectral methods \cite {FRLT09, IPI05, MO92}.
Our formulation with a Neumann to Dirichlet operator can be solved
with boundary element methods. See {\it e.g.} the books of J.C.~N\'ed\'elec \cite{Ne01} or  
O.M.~Faltinsen and A.N. Timokha \cite{FRLT09}. For an explicit implementation 
of  boundary integral  methods, we refer to the work of one of us \cite{DSST05}. 
We observe also that
 equation (\ref{sloshing-classique}) will be modified by the coupling with the rigid movement.

\bigskip \bigskip  \noindent {\bf \large 2) \quad  Sloshing in a body with a rigid movement }  
 
  \noindent
In their book \cite{MR65},  N.~Moiseev and  V.~Rumiantsev study the 
problem of a completely fluid-filled reservoir. They introduce 
special functions to represent the effect of the fluid
on the solid motion. In the present study, the reservoir 
is partially filled and the fluid has a free surface.
Nevertheless, the system is now the rigid body submitted to various forces. 
%
In this section, we derive the evolution equations of momentum and 
kinetic momentum of the solid. We adapt also  the previous section 
in order to describe the fluid movement inside the moving rigid body. 

\smallskip     \monitem   {\bf      Rigid body}  

  \noindent
We consider a rigid moving solid   $\cal S$, of density    $\rho_S$ 
and total mass $ \, m_S $. 
We introduce the center of gravity $ \, \xi  $.   
This solid is submitted to three forces. The first one is the gravity described previously. 
The weight of the solid ${\cal S}$ is then equal to  
$ \,   m_S \, g_0 . \, $  
Secondly a force $R$ at a fixed point  $A$ on the   boundary $\partial {\cal S}$.
We can suppose that this force is a given  function of   time.
Last but not least, the surface  
forces $f$ on the  boundary  $\partial {\cal S}$ 
due to the   internal fluid. 
We introduce a  local referential $\, \varepsilon_j \,$ 
associated to the rigid body and issued 
from Galilean referential  $\, {\rm e}_j . \,$

%

\smallskip   \monitem   {\bf      Infinitesimal motion of the rigid body} 

  \noindent
The linearization hypothesis acts now in a geometrical manner. 
The center of gravity is a function of time    
$ \, \xi \,=\, \xi(t) \,   $  and 
an infinitesimal rotation of angle $ \,  \theta   =  \theta(t) $
allows to write a simple algebraic relation between the vectors  $\, \varepsilon_j \,$ 
and   $\, {\rm e}_j $: 
$ \,\, \varepsilon_j \,=\, {\rm e}_j  \,+\, \theta \times  {\rm e}_j   $. 
Then 
$ \,\, 
 {{{\rm d}\varepsilon_j}\over{{\rm d}t}} \,=\,   {{{\rm d}\theta}\over{{\rm d}t}} 
  \times   \varepsilon_j  \,\, $ for $ \, 1 \leq j \leq 3 $.   
%
The solid velocity field $ \,   u_S(x) \, $ satisfies:
\moneqstar 
 u_S(x) \,=\,  {{{\rm d}\xi}\over{{\rm d}t}}  +  {{{\rm d}\theta}\over{{\rm d}t}} 
   \times   (x- \xi(t) )  \,\,, \qquad  x \in {\cal S} \, .  
\monendstar 
The  kinetic  momentum   $ \,  \sigma_S   \, $ 
and the  tensor of inertia  $  \,\, {\rm I}_S \,\, $  are 
 defined as usual:\br 
%
$ \,  \sigma_S \,\equiv\, \int_{\cal S} \rho_S \,  (x- \xi)    \times   
u_S(x) \, {\rm d}x \, $ 
and 
%
$ \,   {\rm I}_S   \, \smb \, y   \,\equiv\, 
 \int_{\cal S} \rho_S \,  (x- \xi)  \, \times \,  
\big ( y   \times    (x - \xi ) \big)  \, {\rm d}x \, $ 
for $ \, y  \in \R^3 $.  
 For   a rigid body, we have the classical relation:
\moneqstar    
 \sigma_S \,=\, {\rm I}_S \, \smb \,  {{{\rm d}\theta}\over{{\rm d}t}} \, .  
\monendstar  

\smallskip   \monitem   {\bf    Dynamics equations of the rigid body} 

  \noindent 
By  integration of the  classical Newton laws of motion,   
the conservation of momentum takes the form:  
\moneq  \label{impulsion-solide}  
m_S \, {{{\rm d}^2 \xi }\over{{\rm d} t^2}}  \,=\, m_S \, g_0 
\,\,+\,\, R \,\,+\,\,  \int_{\cal \partial S} f \, \,  {\rm d}\gamma \, .
\monend   
The momentum $ \,  {\cal M}_S \, $ of the surface forces 
  relatively to the  center of gravity
is given according to: 
$ \,  {\cal M}_S    \,=\,  
 \int_{\cal \partial S} (x-\xi) \,  \times \, f \,  {\rm d}\gamma  $. 
%
Then the conservation of kinetic momentum takes the form:
\moneq  \label{moment-cinetique}  
 {\rm I}_S   \, \smb \,  {{{\rm d}^2 \theta}\over{{\rm d} t^2}}  \,=\, 
(x_A-\xi)    \times  R \,\,+\,\,  {\cal M}_S     \,.
\monend   

\monitem  
The force $f$ on the boundary of the solid  surface $\, \Sigma \,$ admits the expression 
\moneqstar 
f \,=\,  \left\{ \begin{array}{rcl}
    \displaystyle   p \, n \,,  &&  x \in \Sigma    \\ 
 \displaystyle  \,\, 0  \,,  \,\,  &&  x \in \partial {\cal S} 
\, \setminus \,  \Sigma  \, . 
\end{array}  \right.    \monendstar 
Then: 
$ \, \, \int_{ \partial {\cal S} } f \, {\rm d}\gamma \,= $ 
$ \, \int_{ \Sigma} p \, n  \, {\rm d}\gamma \,= $ 
$ \, \int_{ \partial \Omega } p \, n  \, {\rm d}\gamma \,= $ 
$ \, \int_{  \Omega } \nabla p \, {\rm d} x $.  
Due to the  previous expression  
of the momentum of  pressure forces, we have 
%
%
after an elementary calculus: 
\moneq  \label{moment-forces-solide}  
{\cal M}_S    \,=\,  \, \int_{   \Omega } (x - \xi) \,\times\, \nabla p \, 
 \, {\rm d} x  \,  .
\monend

\smallskip \smallskip   \monitem   {    \bf   About  the fluid irrotationality  hypothesis } 
  
\noindent 
 In the  monograph \cite{Lo64a}, 
  D.~Lomen suppose the   irrotationality for the   motion 
of the   liquid   {\bf relatively} to  the  
motion of the  rigid body.  Then  the velocity field of the liquid 
satisfies the conditions:
\moneqstar   
 u(x) \,=\,  {{{\rm d}\xi}\over{{\rm d}t}}  +  {{{\rm d}\theta}\over{{\rm d}t}} 
  \, \times \, (x- \xi(t) ) \,  + \,v \,\,, \qquad   
   {\rm curl} \, v   \,\equiv\, 0 \,. 
\monendstar
By taking the curl of this relation: 
 $ \,\,   {\rm curl} \, u \,= \,  2 \,  {{{\rm d}\theta}\over{{\rm d}t}}   $ .
%
Consider now the time derivative of the previous relation 
 and the curl  of relation (\ref{Euler-linear}). Then we obtain: 

\smallskip  \centerline  {  $ \displaystyle \, 
  \frac{{\rm d^2}\theta}{{\rm d}t^2}  \, = \,  0  \, $} 

\smallskip  \noindent 
and the hypothesis of irrotationality in the {\bf relative} referential 
done in  \cite{Lo64a}  is physically correct 
{\bf only} if the rotation of solid  referential is {\bf  uniform}  in   time. 
   
    
\smallskip \smallskip   \monitem   {    \bf   Irrotationality in  the Galilean referential }

\noindent
We observe that under an assumption of linearized dynamics, 
if   vorticity $ {\rm curl} \, u \,$ of   liquid measured  
{\bf in the Galilean  referential}
at   initial time is null, then it remains identically null  for all  times: 
\moneqstar 
   {\rm curl} \, u   \,\equiv\, 0 \,, \qquad t \geq 0 
\,, \quad x \in \Omega(t) \,.
\monendstar
%


\noindent 
To prove the previous relation, just take the  curl of the linearized  dynamics
equation  (\ref{Euler-linear}). Then  
 $ \, {{\partial }\over {\partial t}} \big( {\rm curl} \,  u \big) \,=\, 0 \, $ 
%
and the property is established if it is true at  $ \,\,  t=0 .$ 
In the following,  it is assumed that the fluid
is irrotational in the {\bf Galilean}  referential. 
%
In this case, one can find a velocity potential even if the angular velocity 
is an arbitrary function of time.

\smallskip   \monitem   {\bf Stokes - Zhukovsky   vector fields for 
 fluid  potential decomposition  }

\noindent   
A natural question is the incorporation of the  movement of a rigid body 
inside  the  expression of the  potential $\, \varphi \,$ of velocities.
In other words, 
we have put in evidence the very particular role of the rigid movement 
for the determination of the velocity potential. We recall that 
this dynamics is     a six degrees of freedom system 
described  by the two vectors $ \, \xi(t) \,$ 
and  $ \, \theta(t) . \,$
The remaining difficulty concerns the  solid body   velocity field  
  which {\bf is} rotational. 
Following an old idea   due independently  (at our knowledge) to 
G.~Stokes \cite{St1843},  N.~Zhukovsky \cite{Zh1885} and  
 B.~Fraeijs de Veubeke \cite{dV63}, we 
 introduce  a function   $\, \widetilde { \varphi } \,$ such that: 
\moneq \label{pre-SZV}      \left\{ \begin{array}{rcl}
    \displaystyle 
\Delta   \widetilde { \varphi } \,=\, 0    &{\rm in}& \,\, \Omega   \\ 
 \displaystyle  
{{\partial    \widetilde { \varphi } }\over {\partial n}}  \,=\, 
\Big( {{{\rm d}\xi}\over{{\rm d}t}}  \,+\, 
 {{{\rm d}\theta}\over{{\rm d}t}} \,\times\, (x - \xi) \Big) \,\smb\, n  
   & {\rm on} & \,\, \partial \Omega   \, .
\end{array}  \right.    \monend 
Due to {\bf linearity} of the problem (\ref{pre-SZV}), we can decompose 
the vector field $ \,   \widetilde { \varphi} \,$  
under the  form:
\moneq \label{phi-tilda}  
\widetilde { \varphi } \,\equiv\, \alpha  \,\smb\, {{{\rm d}\xi}\over{{\rm d}t}}  
\,+\,  \beta  \,\smb\,  {{{\rm d}\theta}\over{{\rm d}t}} \, .
\monend 
The  vector fields   $ \, \Omega \ni x   \longmapsto   \alpha(x) \in \R^3 \, $ 
for translation  and 
 $  \,  \Omega \ni x  \longmapsto  \beta(x) \in \R^3  \, $ for rotation 
only depend on the three-dimensional geometry of the liquid. 
Observe that $ \,  \alpha(x) \,$ is homogeneous to a length 
and $ \,  \beta(x) \,$ to a surface. 
We call them the  ``Stokes-Zhukovsky  vector fields'' in this contribution, 
as in the reference  \cite{FRLT09}. 

\bigskip 
\centerline { \includegraphics[width=.70 \textwidth]  {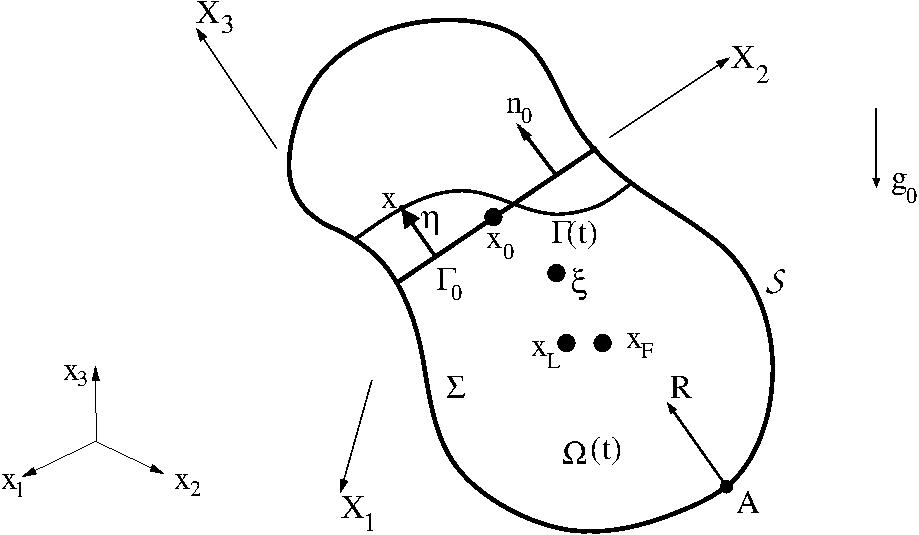} }  

\smallskip \noindent  {\bf Figure 2}. \quad 
General view of the sloshing problem in a six degrees of freedom 
rigid body. The free boundary $ \, \Gamma (t) \, $ 
is issued from the 
equilibrium free boundary at rest  $ \, \Gamma_0 \, $
 with the help of the elongation $ \, \eta . \, $
 The other notations are explained in the corpus 
of the text. 

\smallskip  \newpage   \monitem   {    \bf    Free surface } 
 
  \noindent  
Consider as a reference situation the  fluid at rest relative to the solid.  
Then the  free surface has a given position $ \, \Gamma_0 \,$ 
as presented in Figure~2. 
During sloshing, two processes have now to be taken into account.
First, the   rigid motion of the  surface $ \, \Gamma _0 \,$ 
and secondly the  free displacement $ \, \eta \, n_0 \,$ of the  free boundary
  measured in the relative referential, where 
$\,  n_0 \,$ denotes the   normal direction to $\, \Gamma _0 \,$  
at position $ \, y  \in  \Gamma_0 .\,$ 
%
The local coordinates $ \, X_j \,$ are defined by the relation 
$ \,\, x - x_0 \,=\,  \sum_{j=1}^{3} X_j \, \varepsilon_j  $,  
where $ \, x_0 \, $ is the barycenter  of the free surface $ \, \Gamma_0 \, $ 
defined  in (\ref{centre-surface-libre}). 
We introduce  the center of gravity  $ \, x_L\,$ of the  liquid: 
$ \,\, m_L \,  x_L \, \equiv $ $ \,  \int_{   \Omega }  \rho_L \, x \, {\rm d} x  $,  
which is {\it a priori} a function of   time. 
We introduce also the center of gravity  $ \, x_F\,$ of   the ``frozen fluid''
$\, \Omega_0 \,$ at rest. Note that $ \, \Gamma_0 \,$ is a part of the boundary of 
$\, \Omega_0 $:  
$ \,\, m_L \,  x_F \, \equiv $ $ \,  \int_{   \Omega_0 }  \rho_L \, x  \, {\rm d} x  \, $ 
%
and we refer to Figure~2 for a representation of this point.

\smallskip   \monitem   {\bf    Stokes-Zhukovsky vector fields for translation  }  

  \noindent
We set  
$   \, \alpha \, \equiv \, \sum_{j=1}^3  \alpha_j \, \varepsilon_j  . \,$  
Then the scalar function $ \, \alpha_j(x) \,$ satisfies clearly 
the following Neumann problem for the Laplace equation: 
\moneq \label{pb-alpha}      \left\{ \begin{array}{rcl}
\displaystyle \Delta \alpha_j \,=\, 0    &{\rm in}& \,\, \Omega_0   \\ 
 \displaystyle  
{{\partial  \alpha_j }\over {\partial n}}  \,=\, n_j  
   & {\rm on} & \,\, \partial \Omega_0   \, .
\end{array}  \right.    \monend 
The  problem (\ref{pb-alpha}) has a unique solution 
up to a scalar constant if the domain 
$\, \Omega \,$ is connected. It has  an analytical solution.
We consider the center of gravity $ \,  x_0 \, $ of the linearized free surface 
$ \, \Gamma_0 \,$ according to   (\ref {centre-surface-libre}). Then:  
$ \,\, \alpha_j (x)  \,=\, (x - x_0)  \,\smb\, \varepsilon_j \, $ 
for  $ \,  j = 1, \, 2 ,\, 3  \,$  
if the condition 
$ \,\, \int_{\Gamma_0} \alpha_j  \, {\rm d} \gamma \,=\, 0 $. 
%
holds. Then in consequence, 
\moneqstar 
\nabla \alpha_j  \,=\, \varepsilon_j  \,, \quad j = 1, \, 2 ,\, 3  \,
  \monendstar  

\monitem  
We introduce $ \, \widetilde{\alpha} \, $ by rotating the 
Stokes-Zhukovsky translation vector field 
$ \, \alpha $:  
\moneqstar 
 \widetilde{\alpha} \,\equiv \, ( x - x_0 ) \times \varepsilon_3 \, \, 
= \,\, \alpha \times \varepsilon_3 \,
\monendstar 
%
%
Then: 
$ \,\, \int _{\Gamma_0}  \widetilde{\alpha} \, {\rm d}\gamma = 0  \,\,$ 
and  
$ \,\,  \widetilde{\alpha} \, = \, X_2 \,  \varepsilon_1 \,-\,  X_1 \,  \varepsilon_2  $. 
%
Moreover     $ \,   \,   
\int_{   \Gamma_0 }  \eta \, ( X_2 \, \varepsilon_1  \, - \, 
X_1\, \varepsilon_2 ) \,   {\rm d} \gamma  \,  \,=\,  
\int_{   \Gamma_0 }  \eta \,   \widetilde{\alpha}  \,   {\rm d} \gamma  \, \, $ 
and  
\moneq \label{alpha-tilda-theta} 
 \widetilde{\alpha}  \,\smb \,  \theta  \,= \, 
 ( -X_1 \, \theta_2  \,+\,  X_2 \, \theta_1 )  \, . 
\monend 

\smallskip      \monitem   {\bf   Stokes-Zhukovsky vector fields for rotation   }  

  \noindent 
Analogously to the definition (\ref{pb-alpha}) of  Stokes-Zhukovsky vector fields 
for translation, we set    
$  \, \beta \, \equiv \, \sum_{j=1}^3  \beta_j \, \varepsilon_j  . \,$ 
The scalar function $ \, \beta_j(x) \,$ satisfies the equations:
\moneq \label{pb-beta}      \left\{ \begin{array}{rcl}
\displaystyle \Delta \beta_j \,=\, 0    &{\rm in}& \,\, \Omega_0   \\ 
 \displaystyle  
{{\partial  \beta_j }\over {\partial n}}  \,=\, \big( (x-\xi) \,\times\, n\big)_j 
   & {\rm on} & \,\, \partial \Omega_0   \, .
\end{array}  \right.    \monend 
It is elementary (see {\it e.g.} P.A. Raviart and J.M. Thomas \cite{RT83}) 
 to verify that the   Neumann problem   
(\ref{pb-beta})   is well set  up to an additive constant. 
But, oppositely to     the  Stokes-Zhukovsky vector field  for translation, 
we have {\bf no analytical expression}  for the Stokes-Zhukovsky
functions  $ \,  \beta_j \,$  for rotation. 
%
Nevertheless, 
the following relations show that beautiful algebra can be developed for
the Stokes-Zhukovsky vector fields. They are proven in detail in  \cite{DS14}:
\moneq \label{tech-prop-i}  
\rho_L \, \int_{\Omega} \nabla \alpha_j \, {\rm d}x = m_L \, \varepsilon_j  \,,
\quad 
\rho_L \, \int_{\Omega} \nabla \beta_j \, {\rm d}x =  m_L \, \varepsilon_j  \,\times\, (x_F-\xi) 
\,, \quad j = 1, \, 2 ,\, 3  \, .  
 \monend

\smallskip   \newpage \monitem   {\bf      Liquid inertial tensor   }  

\noindent 
With  B. Fraeijs de Veubeke \cite{dV63}, 
we introduce the so-called ``liquid inertial tensor''  
 $ \, {\rm I}_{\ell} \,$ 
defined according to 
\moneq \label{inertie-liquide}  
\big(  {\rm I}_{\ell} \big) _{j\, k}  \,\equiv\, \rho_L \, \int_{\Omega} 
 \nabla \beta_j   \,\smb\,  \nabla \beta_k   \, {\rm d}x   \, . 
  \monend    
We have the complementary results, proven also in \cite{DS14}:
\moneq \label{tech-prop-iii}  
\left\{ \begin{array}{rcl} \displaystyle  
\rho_L \, \int_{\Omega} (x-\xi) \,\times\, \nabla \alpha_j \, {\rm d}x  & = &  
m_L \, (x_F-\xi) \,\times\, \varepsilon_j   \,, \quad j = 1, \, 2 ,\, 3  \,, 
 \\ \displaystyle   & & \vspace{-.5cm}  ~  \\ \displaystyle  
\rho_L \, \int_{\Omega}  (x-\xi) \,\times\, \nabla \beta_j \, {\rm d}x   & = &  
{\rm I}_{\ell} \,\smb\, \varepsilon_j   \,   
 \,, \quad j = 1, \, 2 ,\, 3  \,. 
\end{array} \right. \monend
%
%
Moreover, the liquid inertial tensor $ \,  {\rm I}_{\ell}  \, $ 
defined in (\ref{inertie-liquide}) is positive definite. We  have 
$ \, \big ( \theta \,, \,   {\rm I}_{\ell} \,\smb\, \theta \big)  \,=  $ 
$ \, \rho_L  \, \int_\Omega \mid \! \nabla ( \beta \smb \theta ) \! \mid^2 \, {\rm d}x \,\geq \,\, 0 \,$
for $ \,  \theta \in \R^3 $. 
Moreover,  if $ \, ( \theta \,, \,   {\rm I}_{\ell} \,\smb\, \theta  ) = 0 $, then 
$\, \theta = 0 \,$ in $\, \R^3 $.


\smallskip    \monitem   {\bf      Decomposition of the fluid velocity potential  }  

\noindent 
The fluid velocity potential $ \, \varphi \,$ 
satisfies a continuity condition across the solid interface $ \,  \Sigma (t) \,$
due to the non-penetration of the fluid inside the solid: 
\moneq    \label{neumann-solide}  
    {{\partial \varphi}\over{\partial n}} \,= \, 
 \Big( {{{\rm d}\xi}\over{{\rm d}t}} \,+\,  {{{\rm d}\theta}\over{{\rm d}t}} \, \times \, 
(x - \xi) \,  \Big) \, \smb \, n \,, \qquad x \in \Sigma . \,
\monend 
Therefore we subtract to $ \, \varphi \,$  the potential 
$ \, \widetilde { \varphi }  \,$ introduced in (\ref{phi-tilda}) and 
the difference satisfies a  homogeneous boundary condition on  the solid interface.
In an analogous way,  Proposition~1 can be derived in the relative referential. 
Then the  fluid velocity potential  $ \, \varphi \,$ can be decomposed according to 
\moneq \label{phi-psi}       
 \varphi  \,\equiv\, \alpha  \,\smb\,  {{{\rm d}\xi}\over{{\rm d}t}} \,+\, 
 \beta  \,\smb\,  {{{\rm d}\theta}\over{{\rm d}t}}   \,+\, 
  {{\partial \psi }\over{\partial t }} \, . 
\monend 
The free surface potential $ \, \psi \,$ introduced in (\ref{phi-psi}) 
still satisfies  the relations (\ref{pb-psi}): 
\moneqstar     \left\{ \begin{array}{rcl}
\displaystyle \Delta   \psi &\,=\,&  0    \qquad {\rm in} \,\,   \Omega   \\ 
 \displaystyle     {{\partial \psi}\over{\partial n}}  &\,=\, & 
 \left\{ \begin{array}{rcl}
0   & {\rm on} &   \Sigma   \\
 \eta   & {\rm on} &     \Gamma_0   \, . 
\end{array}   \right. \end{array}    \right.    \monendstar
Due to the definition (\ref{def-W})  of the Neumann to Dirichlet operator, 
this last expression admits the compact form
\moneq \label{eta-to-psi}       
\psi \,=\, W \smb \,  \eta  \, \quad {\rm on} \,\, \Gamma_0 \, . 
 \monend

\monitem 
We introduce   the    position  $ \, \ell_0 \,$ of the center of gravity of the fluid
  relatively to the solid center of gravity: 
\moneq \label{f-zero}   
 \ell_0  \,\equiv \, x_F - \xi  \, . 
\monend 
We observe that this vector is linked to the solid and we have 
in particular
$ \,  {\rm d} \ell_0 =   {\rm d}  \theta \times  \ell_0 \, . $   
%
We have the following relations, 
with  $ \, \, \eta \in F^{1/2} (\Gamma_0)  \, $ and 
$ \, \alpha ,\, $   $ \, \beta \, $ defined in
 (\ref{pb-alpha})(\ref{pb-beta}) and $ \, \psi \,$ by the relation  (\ref{pb-psi}). 
The proofs are detailed in our report \cite{DS14}. 
\moneq \label{moments-psi-eta}  
\left\{ \begin{array}{rcl} \displaystyle  
\rho_L \, \int_{\Omega} \nabla \bigg(  {{\partial^2 \psi}\over{\partial t^2}} 
\bigg)   \, {\rm d}x    & = &   \displaystyle 
 \rho_L \, \int_{\Gamma_0 }  {{\partial^2 \eta}\over{\partial t^2}} \,  
  (X_1 \, \varepsilon_1 \,+\, X_2 \, \varepsilon_2 )   \, {\rm d}\gamma   
 \\ \displaystyle   & & \vspace{-.5cm}  ~  \\ \displaystyle 
\rho_L \, \int_{\Omega} (x-\xi) \,\times \,\nabla \bigg( {{\partial^2 \psi}\over{\partial t^2}} 
\bigg)   \, {\rm d}x  & = &  \displaystyle   \rho_L \, \int_{\Gamma_0 } 
 {{\partial^2 \eta}\over{\partial t^2}} \, \beta   \, {\rm d}\gamma   \, . 
\end{array} \right. \monend
%
%

\bigskip   \noindent $\bullet$ \quad    {    \bf Proposition 4.   
Pressure continuity      across the free surface     } 
%

 \noindent
The continuity of pressure on the free boundary  $\, \Gamma \, $ takes the following linearized
form: 
\moneq  \label{pression-surface-libre-bis}    
 {{\partial \varphi}\over{\partial t }} \,+\,
g \, \big( \widetilde{\alpha}  \,\smb \,  \theta \,+\, \eta \big) 
 \, = \,  0 \,, \qquad x \in \Gamma(t)  \, .
\monend 
In an equivalent way with the potential $ \, \psi \,$ introduced in 
(\ref{phi-psi}):
\moneq  \label{pression-surface-libre-ter}   
{{\partial^2 \psi}\over{\partial t^2}}  \,+\, 
\alpha \,\smb\,   {{{\rm d}\xi}^2\over{{\rm d}t^2}} \,+\, 
\beta \,\smb\,    {{{\rm d}^2 \theta}\over{{\rm d}t^2}}\,+\, 
g \, \big( \widetilde{\alpha}  \,\smb \,  \theta \,+\, \eta \big) \,=\, 0 
\qquad {\rm on } \,\,\, \Gamma_0 \,. 
\monend   
Compared to the relation (\ref{pression-surface-libre-ini}), the new term 
$ \, g \, \widetilde{\alpha}  \,\smb \,  \theta \,$ expresses  the 
action of the external gravity field  driving 
the fluid due to the rotational  movement of the solid. 
In the stationary case, the local displacement $ \, \eta \,$ of the free boundary
compensates exactly the rotational  displacement of the solid. 

\bigskip   \noindent $\bullet$ \quad    {    \bf  Proof of Proposition 4. }  

\noindent 
The proof is a small variation of the one proposed for Proposition~2. 
On the free surface $ \,  \Gamma ,\,$ the    
continuity of the  stress tensor   can be written $ \, p = 0 \,$ 
as previously.  
We choose  the   point $\, P  \,$ for the  Bernoulli equation  (\ref{bernoulli})
 on the  frozen free surface $\, \Gamma_0\,$  equal to the center 
$\, x_0 \, $ introduced in  (\ref{centre-surface-libre}).  
Then 
$ \,\, x - x_0  \,=\, X_1 \, \varepsilon_1 \,+\,  X_2 \, \varepsilon_2 \,+\, 
 \eta   \,  \,  \varepsilon_3  \,+\, {\rm O} (\vert \eta\vert^2)  $ .
Due to Bernoulli theorem  (\ref{bernoulli}) and continuity (\ref{pression-surface-libre-zero})
of the pressure on  $\, \Gamma $, we deduce the following relation (\ref{pression-surface-libre-initial})
on the free surface. 
In order to show   the angular displacement of the solid, 
we have the following calculus:

\smallskip   $ \displaystyle \,
-\, g_0 \,\smb \, (x - x_0)  \,=\, g\,  {\rm e}_3  \,\smb \, \Big(  
 X_1 \,  ( {\rm e}_1  \,+\, \theta \,\times\,  {\rm e}_1 )  \,+\, 
 X_2 \,  ( {\rm e}_2  \,+\, \theta \,\times\,  {\rm e}_2 )  
\,+\, \eta \, \varepsilon_3 \Big) \,,  $

\smallskip  \qquad \qquad \qquad  $ \,\,\,\, \,  \displaystyle \, = \, 
g \, ( -X_1 \, \theta_2  \,+\,  X_2 \, \theta_1 \,+\, \eta )  
\,+\, {\rm O} (\vert \eta\vert^2) \, , $

\smallskip \noindent  
and the condition   (\ref{pression-surface-libre-initial})   of 
pressure continuity  on  the free surface is expressed by
\moneqstar     
 {{\partial \varphi}\over{\partial t }} \,+\,
g \, ( -X_1 \, \theta_2  \,+\,  X_2 \, \theta_1 \,+\, \eta )  
\,+\, {\rm O} (\vert \eta\vert^2)  \, = \,  0 \,, \qquad x \in \Gamma(t) \,,\,  
\monendstar
which is exactly relation (\ref{pression-surface-libre-bis}) due to the relation 
(\ref{alpha-tilda-theta}): $\,   \widetilde{\alpha}  \,\smb \,  \theta 
 \,= \,  ( -X_1 \, \theta_2  \,+\,  X_2 \, \theta_1 )  .$  
%
%
If we determine    the velocity potential $ \, \varphi \,$ 
according to the left hand side of the relation  (\ref{phi-psi}), 
the continuity of the pressure field   
(\ref{pression-surface-libre-zero})
   across the  free surface takes exactly the form 
(\ref{pression-surface-libre-ter}) and the proof is completed. 
\hfill  $ \square $ 

\bigskip    \noindent $\bullet$ \quad    {    \bf Proposition 5.  
Conservation of  the solid momentum } 

\noindent 
The conservation of the solid momentum conducts to the coupled relation:
\moneq \label{impulsion-solide-4}  
( m_S + m_L) \, {{{\rm d}^2\xi}\over{{\rm d}t^2}}  \,-\,  m_L \, \ell_0 \times  
  {{{\rm d}^2 \theta}\over{{\rm d}t^2}}    \,+\, 
 \rho_L \, \int_{\Gamma_0 } \alpha \, 
  {{\partial^2 \eta}\over{\partial t^2}} \,  {\rm d}\gamma 
 \,=\,  ( m_S + m_L)  \, g_0  \,+\, R  \,. 
 \monend

\bigskip   \noindent $\bullet$ \quad    {    \bf  Proof of Proposition 5. }  

\noindent 
Due to the linearized Euler equations (\ref{Euler-linear}), the pressure field   action can be stated as 
\moneqstar 
 \nabla p  \,=\,   \rho_L \, \Big[  g_0 \,-\, \nabla 
\Big(   {{\partial \varphi}\over{\partial t }} \Big) \Big] \,, \qquad x \in \Omega \,,
\monendstar 
the conservation of momentum of the solid can be written as: 
\moneq \label{conservation-impulsion-1}  
m_S \, {{{\rm d}^2\xi}\over{{\rm d}t^2}}  \,=\, ( m_S + m_L)  \, g_0 
\,+\, R \,-\,  
\int_{  \Omega }   \rho_L \, \nabla 
\Big(   {{\partial \varphi}\over{\partial t }} \Big)  \, {\rm d} x  \, .
\monend 
\noindent 
With the help of the Stokes-Zhukovsky vector fields, we 
can express the last term in the right hand side of  (\ref{conservation-impulsion-1})
with the free surface potential $ \, \psi $:

\smallskip \smallskip \noindent    $ \displaystyle   \rho_L \, \int_{\Omega} \,
 \nabla  \Big( {{\partial \varphi}\over{\partial t}}  \Big) \, {\rm d}x   \,=\,  \rho_L \, 
 \int_{\Omega} \, \Big[ \,  \sum_j  \Big( \nabla \alpha_j \, {{{\rm d}^2 \xi_j }\over{{\rm d}t^2}}  
\,+\,  \nabla \beta_j \,   {{{\rm d}^2 \theta_j }\over{{\rm d}t^2}}  \Big) \, \,+\, 
\nabla  \Big(  {{\partial^2 \psi}\over{\partial t^2}}   \Big)   \, \Big] \, {\rm d}x  $ 
\hfill {\it c.f.} (\ref{phi-psi}) 

 \smallskip   \noindent     $ \displaystyle 
= \, m_L \,  {{{\rm d}^2 \xi }\over{{\rm d}t^2}}  \,+\, m_L \, 
  {{{\rm d}^2 \theta }\over{{\rm d}t^2}} \,\times \, \big( x_F - \xi \big) 
 \,+\,   \rho_L \, \int_{\Omega} \,  {{\partial^2 \eta}\over{\partial t^2}} 
\, ( X_1 \, \epsilon_1 +  X_2 \, \epsilon_2 ) \,  {\rm d}\gamma $ 

 \smallskip   \noindent  
due to  (\ref{tech-prop-i}),  (\ref{tech-prop-iii}) 
and     (\ref{moments-psi-eta}).   
Then the conservation of impulsion of the solid  (\ref{conservation-impulsion-1}) 
takes the form:
\moneqstar 
\left\{ \begin{array}{c}
\displaystyle ( m_S + m_L) \, {{{\rm d}^2\xi}\over{{\rm d}t^2}}  \,+\, 
 m_L \,  {{{\rm d}^2 \theta}\over{{\rm d}t^2}}  \,\times \, (x_F - \xi )  \,+\, 
 \rho_L \, \int_{\Gamma_0 } 
  {{\partial^2 \eta}\over{\partial t^2}} \, (X_1 \, \varepsilon_1 \,+\, X_2 \, \varepsilon_2 ) 
  \, {\rm d}\gamma  \,=\,  \\
\displaystyle  \hfill \qquad \qquad \qquad \,=\,   ( m_S + m_L)  \, g_0  \,+\, R   \, . 
\end{array} \right. \monendstar 
%
\smallskip \noindent 
Then due to (\ref{f-zero}), 
the equations  (\ref{impulsion-solide}) 
for the conservation of impulsion of the   solid
takes  the following expression (\ref{impulsion-solide-4}) for the  coupled problem.
The proof is completed.
\hfill  $ \square $ 

%
\bigskip   \noindent $\bullet$ \quad    {    \bf Proposition 6. 
 Conservation of  the solid angular momentum } 

\noindent 
The conservation of the solid angular momentum conducts to the coupled relation
\moneq \label{moment-cinetique-4}    \left\{ \begin{array}{c}
\displaystyle  
m_L \, \ell_0 \times  {{{\rm d}^2\xi}\over{{\rm d}t^2}}   \,+\,
( {\rm I}_S + {\rm I}_{\ell} )   \, \smb \,  {{{\rm d}^2 \theta}\over{{\rm d}t^2}} \,+\, 
 \rho_L \, \int_{\Gamma_0 }  \beta \,
 {{\partial^2 \eta}\over{\partial t^2}} \,   {\rm d}\gamma     \,\, \,+\,   
 \rho_L \,  g \,   \int_{   \Gamma_0 }   \widetilde{\alpha}  \, \eta \,\,
   {\rm d} \gamma  \,  \,=\,  
\hfill \\   \displaystyle   \hfill  \hfill    \,=\, 
   m_L \, \ell_0  \times  g_0  \,\,+\,\,  (x_A-\xi)   \times   R \, . 
\end{array} \right. \monend 
%
The term $ \, m_L \, \ell_0 \times  {{{\rm d}^2\xi}\over{{\rm d}t^2}}   \, $
of dynamic evolution of the kinetic momentum in the relation (\ref{moment-cinetique-4})
is due to the non-coincidence of the center of gravity $ \, \xi \, $ of the solid
and the frozen center of gravity $ \, x_F \, $ of the frozen liquid (see Figure~2).  
With the help of (\ref{moments-psi-eta}), the kinetic momentum 
of the fluid relative to the center of gravity of the solid 
is represented by the term  $ \, \rho_L \, \int_{\Gamma_0 }  \beta \,\smb \, 
 {{\partial^2 \eta}\over{\partial t^2}} \,   {\rm d}\gamma . \,$ 
Last but not least, the restoring torque  of the gravity field due to the 
weight of the liquid displaced by the movement of the free boundary is described by 
$ \, \rho_L \,  g \,   \int_{   \Gamma_0 }   \widetilde{\alpha}  \, \eta \,  {\rm d} \gamma .$ 
We can interpret the elongation of the free surface by a continuous distribution 
 of small harmonic oscillators that have an impact on the global conservation of
the solid angular momentum. 

\bigskip   \noindent $\bullet$ \quad    {    \bf  Proof of Proposition 6. }  

%

  \noindent 
We explain  in the following how the center of gravity  $ \, x_L \,$ of the  liquid 
depends on the position $ \, \eta \,$
relative to the free boundary.
The conservation of  kinetic momentum (\ref{moment-cinetique})
 takes now the form:
\moneqstar 
 {\rm I}_S   \, \smb \,  {{{\rm d}^2 \theta}\over{{\rm d} t^2}}  \,=\, 
(x_A-\xi)    \times  R  \,\,+ \,\, m_L \, (x_L - \xi)  \times  g_0  
 \,-\,  \int_{   \Omega }  \rho_L \, (x - \xi)  \,\times \,   \nabla 
\Big(   {{\partial \varphi}\over{\partial t }} \Big)  \, {\rm d} x . \, 
\monendstar 
We consider also the center of gravity  $ \, x_F \,$ of   the ``frozen fluid'' and 
we denote by $\, X_3^0 \,$ the vertical coordinate of the  frozen free surface $\,  \Gamma_0 .\,$
Relatively to the   rigid referential, we have the following calculus:

\bigskip  \noindent  $ \displaystyle \, 
m_L \, x_L \, = \,  \int_{   \Gamma_0 } {\rm d}X_1 \,  {\rm d}X_2 \, 
\bigg(  \int_{X_{3 \, {\rm min}}}^{X_3^0} \rho_L \, x \,  {\rm d}X_3 \, + \, 
  \int_{X_3^0}^{X_3^0 + \eta} \rho_L \, x \,  {\rm d}X_3  \bigg) \, $

\smallskip  \noindent \qquad  \quad   $ \displaystyle \, 
 \, = \, m_L \, x_F   \, + \,  \int_{   \Gamma_0 } {\rm d}X_1 \,  {\rm d}X_2 \, 
 \int_{0}^{\displaystyle \eta} \, \rho_L \,
 \begin{pmatrix} X_1 \cr X_2 \cr  X_3 \cr \end{pmatrix}  \,  {\rm d}X_3 \,
 \, = \, m_L \, x_F   \, + \,  \int_{   \Gamma_0 } \, \rho_L \, 
 \begin{pmatrix} X_1 \, \eta \cr X_2  \, \eta \cr {1\over2} \eta^2  \end{pmatrix}  \, 
 {\rm d} \gamma  \,.   $   

\smallskip \noindent 
At   first order:
$ \,\, m_L \,  x_L   \times  g_0  \, = $ 
$  \, m_L \, x_F  \times  g_0    + 
 \rho_L \,  g \,   \int_{   \Gamma_0 }  \eta \, ( -X_2 \, \varepsilon_1  + 
X_1\, \varepsilon_2 ) \,   {\rm d} \gamma  $.
%

  \monitem  In consequence the conservation of   kinetic momentum can be written under the form:
\moneq \label{conservation-moment-cinetique}      \left\{ \begin{array}{rcl}
    \displaystyle {\rm I}_S   \, \smb \,  {{{\rm d}^2 \theta }\over{{\rm d} t^2}}  \,&=&\, 
(x_A-\xi)   \times  R  \,\,+ \,\, m_L \, (x_F - \xi) \times g_0     \\ 
 \displaystyle  &+& \displaystyle  \,   \rho_L \,  g \,   
\int_{   \Gamma_0 }  \eta \, ( -X_2 \, \varepsilon_1  \, \, + \, \, 
X_1\, \varepsilon_2 ) \,   {\rm d} \gamma  \,  
 \,-\,  \int_{   \Omega }  \rho_L \, (x - \xi)  \times    \nabla 
\Big(   {{\partial \varphi}\over{\partial t }} \Big)  \, {\rm d} x . \, 
\end{array}  \right.    \monend 
In an analogous way, the  
last term of the  right hand side of  (\ref{conservation-moment-cinetique})
can be developed with the help of the decomposition (\ref{phi-psi}):

\smallskip \smallskip \noindent    $ \displaystyle  
 \rho_L \, \int_{\Omega} \,  \big( x - \xi \big) \,\times \,  \nabla  
 \Big( {{\partial \varphi}\over{\partial t}}  \Big)  \, {\rm d}x   \,=\, $

\smallskip \noindent    $ \displaystyle   \qquad   \qquad   \qquad  \qquad 
= \,  \rho_L \, \int_{\Omega} \,  \big( x - \xi \big) \,\times \, 
 \Big[ \,  \sum_j  \Big( \nabla \alpha_j \, {{{\rm d}^2 \xi_j }\over{{\rm d}t^2}}  
\,+\,  \nabla \beta_j \,   {{{\rm d}^2 \theta_j }\over{{\rm d}t^2}}  \Big) \, \,+\, 
\nabla  \Big(  {{\partial^2 \psi}\over{\partial t^2}}   \Big)   \, \Big] \, {\rm d}x  $ 

\smallskip \noindent    $ \displaystyle   \qquad   \qquad   \qquad  \qquad 
= \, m_L \,  \big( x_F - \xi \big)  \,\times \, 
 {{{\rm d}^2 \xi}\over{{\rm d}t^2}}  \,+\, 
{\rm I}_\ell \, \smb \,   {{{\rm d}^2 \theta}\over{{\rm d}t^2}}  \,+\, 
 \rho_L \, \int_{\Omega} \,  {{\partial^2 \eta}\over{\partial t^2}}  \,
\beta  \,  {\rm d}\gamma $
 
 \smallskip   \noindent  
due to  (\ref{tech-prop-i}),  (\ref{tech-prop-iii}) and     (\ref{moments-psi-eta}).
In consequence, the 
motion (\ref{conservation-moment-cinetique})         
of the   solid around its center of gravity can be written as: 
\moneqstar 
\left\{ \begin{array}{c} \displaystyle  
( {\rm I}_S + {\rm I}_{\ell} )   \, \smb \,  {{{\rm d}^2 \theta}\over{{\rm d}t^2}} \,+\, 
m_L \, (x_F-\xi)  \,\times \, {{{\rm d}^2\xi}\over{{\rm d}t^2}}  \,+\, 
 \rho_L \, \int_{\Gamma_0 } 
 {{\partial^2 \eta}\over{\partial t^2}} \, \beta   \, {\rm d}\gamma   \,+\, 
\hfill \\   \displaystyle  \qquad  \hfill \,+\, 
 \rho_L \,  g \,   \int_{   \Gamma_0 }  \eta \, ( X_2 \, \varepsilon_1  \, - \, 
X_1\, \varepsilon_2 ) \,   {\rm d} \gamma  \,  \,=\, 
  m_L \, (x_F - \xi) \,\times \, g_0  \,+\, (x_A-\xi) \,  \times \, R \, . 
\end{array} \right. \monendstar 
and due to (\ref{f-zero}), 
this is exactly the relation (\ref{moment-cinetique-4}) and the proof is completed.
\hfill  $ \square $ 

\smallskip   \monitem   {    \bf    Towards a synthetic formulation }  

   \noindent 
With the help of the relation (\ref{eta-to-psi}) on the boundary $ \, \Gamma_0 $, 
the continuity (\ref{pression-surface-libre-ter}) of the pressure field 
across the  free surface   is simply written as:
\moneq \label{continuite-pression-4}    
\rho_L \, W  \,\smb\,  {{\partial^2 \eta}\over{\partial t^2}}  \,+\,  
\rho_L \, \alpha \,\smb\,   {{{\rm d}^2\xi}\over{{\rm d}t^2}} \,+\, 
\rho_L \, \beta \,\smb\,    {{{\rm d}^2 \theta}\over{{\rm d}t^2}}\,+\, 
\rho_L \, g \, ( \widetilde{\alpha}  \,\smb \,
 \theta    \,+\, \eta )  \,=\, 0  
\quad {\rm on } \,\,\, \Gamma_0 \,. 
\monend 
The term $ \, \rho_L \, g \,  \widetilde{\alpha}  \,\smb \,   \theta  \,$
is due to the weight of the fluid working in the rigid movement associated to 
the solid rotation. 
%
The coupled problem (\ref{impulsion-solide-4})  (\ref{moment-cinetique-4}) 
(\ref{continuite-pression-4})     is now              
formulated in an attractive  mathematical point of view. 
The unknown is composed of  the triple $\, (  \xi(t) , \,  \theta(t) ,\, \eta(t) ) ,\, $
with  $ \, \xi(t) \in \R^3 ,\, $ 
$ \, \theta(t) \in \R^3 ,\, $ 
$ \, \eta(t) \in F^{1/2} (\Gamma_0)  \, $ 
and the three equations (\ref{impulsion-solide-4})  (\ref{moment-cinetique-4}) 
(\ref{continuite-pression-4}) 
are considered  in  $ \, \R^3 , \, $  $ \, \R^3  \, $ and on $ \,\Gamma_0 \,$
respectively.
The mathematical difficulty is due to the term 
$\, W \, \smb \, {{\partial ^2 \eta}\over{\partial t ^2}} \,$ 
because $ \, W \,$ is an integral operator. 

\bigskip \bigskip   \newpage    \noindent {\bf \large 3) \quad  Coupled   system structure   }

\monitem
We are now in position to aggregate the previous results. 
The solid movement  is a six degrees of freedom motion described 
 by the  velocity 
$ \,   {{\rm d \xi}\over{{\rm d}t}} \, $ 
of its  center of gravity and  its instantaneous rotation 
$ \,   {{{\rm d}\theta}\over{{\rm d}t}}  . \,$ 
The motion of the  solid around its center of gravity has been obtained in  relation
(\ref{conservation-moment-cinetique}). The  two equations 
(\ref{conservation-impulsion-1}) and (\ref{conservation-moment-cinetique})
admit as a source term
the gradient of the velocity potential.
The partial differential equation that governs this potential is  simply 
the incompressibility of the  liquid, expressed by the Laplace equation.
The boundary conditions are the 
non-penetration  (\ref{neumann-solide}) of the  fluid inside the  solid, 
the normal  movement  (\ref{pb-psi})  
of the  fluid relatively  to the  free surface 
and the continuity (\ref{pression-surface-libre-zero})
of the pressure field across the  free surface  
 expressed by (\ref{pression-surface-libre-ter}).


\smallskip   \monitem   {    \bf    Energy  conservation  } 
 
  \noindent  

\smallskip \noindent 
We can now consider the three  terms of  the  total energy:  
the uncoupled kinetic energy 
\moneqstar 
T \, \equiv \, {1\over2}  \, {{{\rm d}\xi}\over{{\rm d}t}} \,\smb\, m_S 
\, {{{\rm d}\xi}\over{{\rm d}t}}   \,+\,  
{1\over2}  \,  {{{\rm d}\theta}\over{{\rm d}t}} \,  {\rm I}_S   \, \smb \, 
  {{{\rm d}\theta}\over{{\rm d}t}}  \,+\,   
{1\over2}  \,  \int_{  \Omega }   \rho_L \, \vert \nabla \varphi \vert^2 
 \, {\rm d} x  \, ,  
  \monendstar  
the energy of interaction with gravity 
\moneqstar 
U  \, \equiv \,  {1\over2}  \,   \rho_L \, g \,  \int_{  \Gamma_0 } 
\vert \eta \vert ^2  \, {\rm d} \gamma  
\,+\,  \rho_L \, g \,   \int_{  \Gamma_0 }  \eta \, ( X_2 \, \theta_1  \,-\,
 X_1 \, \theta_2 )  \, {\rm d} \gamma    
  \monendstar   
and the gravity potential 
 $ \,\, V \, \equiv \, - m_S \, g_0  \,\smb\, \xi  \,-\,  m_L \, g_0 \,\smb\, x_F  $.  
%
With kinetic energy $T$, 
energy of interaction with gravity $U$ and gravity potential $V$
defined previously 
respectively, we have the following detailed expressions:
\moneq \label{kinetic-1011} \left\{ \begin{array}{c} \displaystyle  
T \,= \, {1\over2} (m_S + m_L) \, \vert {{{\rm d}\xi}\over{{\rm d}t}} \vert^2 
\,+\,  {1\over2} \Big(  {{{\rm d}\theta}\over{{\rm d}t}} ,\, 
(I_S + I_\ell ) \,   {{{\rm d}\theta}\over{{\rm d}t}}  \Big) 
\,+\,  {{\rho_L}\over{2}} \, \int_{\Gamma_0}   \Big(
 {{\partial \eta }\over{\partial t}} ,\, W \smb  {{\partial \eta }\over{\partial t}}
  \Big) \, {\rm d}\gamma   
 \\ \displaystyle   \vspace{-.5cm}  ~  \\ \displaystyle  
+ \, m_L  \, \Big(  \ell_0 , \,  {{{\rm d}\xi}\over{{\rm d}t}} ,\, 
  {{{\rm d}\theta}\over{{\rm d}t}}  \Big) + \rho_L \, 
 \int_{\Gamma_0}   \Big( \alpha \,  {{{\rm d}\xi}\over{{\rm d}t}} + 
\beta \,   {{{\rm d}\theta}\over{{\rm d}t}}  \Big) \, 
  {{\partial \eta }\over{\partial t}} \,  {\rm d}\gamma
\end{array} \right. \monend 
\moneq \label{poten-1011} 
U  \,= \, {1\over2} \rho_L \, g \,  \int_{\Gamma_0}  \eta^2 \,  {\rm d}\gamma
\,+\,  \rho_L \, g \,  \int_{\Gamma_0} ( \widetilde{\alpha} \, \smb \, \theta ) \, \eta  
 \,  {\rm d}\gamma
  \monend 
\moneq \label{exte-1011} 
V  \,= \, -  (m_S + m_L) \, \, g_0 \,  \smb \, \xi \,-\,  
m_L \, g_0 \,  \smb \, \ell_0  \, . 
  \monend 
We recognize the kinetic energy of the solid with the translation and
rotation  decoupled terms 
$ {1\over2} (m_S + m_L) \, \vert {{{\rm d}\xi}\over{{\rm d}t}} \vert^2 $, 
$  {1\over2} \big(  {{{\rm d}\theta}\over{{\rm d}t}}  ,\, 
(I_S + I_\ell ) \,   {{{\rm d}\theta}\over{{\rm d}t}}  \big) $, 
the coupling between translation and rotation 
$ m_L  \big(  \ell_0 , \,  {{{\rm d}\xi}\over{{\rm d}t}} ,\, 
  {{{\rm d}\theta}\over{{\rm d}t}}  \big)$, 
the kinetic energy of the free surface 
$  {{\rho_L}\over{2}} \, \int_{\Gamma_0}   \big(
 {{\partial \eta }\over{\partial t}} ,\, W \smb  {{\partial \eta }\over{\partial t}}
  \big) \, {\rm d}\gamma $ and the coupling 
$  \, \rho_L \, 
 \int_{\Gamma_0}   \big( \alpha \,  {{{\rm d}\xi}\over{{\rm d}t}} + 
\beta \,   {{{\rm d}\theta}\over{{\rm d}t}}  \big) \, 
  {{\partial \eta }\over{\partial t}} \,  {\rm d}\gamma \, $ 
between the solid movement and 
the free boundary.

\bigskip   \noindent $\bullet$ \quad    {    \bf Proposition 7. 
 Energy conservation  } 

\noindent
Due to the lack of knowledge concerning the external force $R$, the
conservation of energy takes the following form, 
with   the previous notations: 
\moneq \label{conservation-energie}  
{{\rm d}\over{{\rm d}t}} \big( T \,+\, U \,+\,V \big) \,=\, R  \,\smb\, u_A \,.
  \monend   
The proof is detailed in  \cite{DS14}.  

\bigskip    \monitem   {    \bf    Operator matrices   } 

\noindent
We consider now the global vector $ \, q(t) \, $ according to: 
\moneq \label{global-vector}   
q \,\equiv \, \big(  \eta \,, \xi  ,\, \theta  \big)^{\displaystyle \rm t} \, . 
\monend 
Remark that when  $ \, t \geq 0 , \,$ $\, q(t) \,$ belongs to the functional 
space $ \,  F^{1/2} (\Gamma_0)  \times \R^3 \times  \R^3  ,\,$
an infinite dimensional vector space denoted by 
$ \, Q_0 (\Omega, \, {\cal S}) \,$ in the following:
\moneq \label{espace-Q0}   
 Q_0 (\Omega, \, {\cal S}) \,\equiv \,  
 F^{1/2} (\Gamma_0)  \times \R^3 \times  \R^3 \, . 
\monend
With this notation, 
  the {\bf interaction} between  the liquid $ \, \Omega \,$ 
and the solid $\,{\cal S} $,    through the free boundary $ \, \Gamma_0 $,    
is defined through    global    operator  matrices     $ \, M \,$ and $ \, K . \, $ 
The mass matrix    $ \, M  \,  $ is defined according to:  
\moneq \label{matrice-M} 
M = \begin{pmatrix}    \rho_L \, W  \smb &  \rho_L \, \alpha\,\smb &   \rho_L \,  \beta \,\smb  \\ 
 \displaystyle \,  \rho_L \,   \int_{\Gamma_0 }  {\rm d}\gamma  \, \alpha  \,\smb \, &
m_S + m_L & - m_L \, \ell_0 \times  \smb  \\
 \rho_L \,  \displaystyle \,  \int_{\Gamma_0 }  {\rm d}\gamma  \, \beta  \,\smb \, & 
 \,\, m_L \, \ell_0 \times  \smb \,\, & {\rm I}_S + {\rm I}_{\ell}

\end{pmatrix} \, .  
\monend 
Remark that this matrix is composed by operators. In particular the 
operator $ \, W \, $ at the position $\, ( 1 , \,  1 ) \, $  
is defined in (\ref{def-W}). 
Moreover, if $ \, q \in  Q_0 (\Omega, \, {\cal S}) ,\, $ 
$ \, M \, \smb \, q \in  Q_0 (\Omega, \, {\cal S}) \, $ 
and $\, M \,$  is an operator $\,  Q_0 (\Omega, \, {\cal S}) 
\longrightarrow  Q_0 (\Omega, \, {\cal S}) . \, $ 
In an analogous way, we define the  global rigidity    matrix $ \, K   $:
\moneq \label{matrice-K}   
K = \begin{pmatrix}   
 \rho_L \, g & 0 & \rho_L \, g \,  \widetilde{\alpha} \,\smb  \\ 
0  & 0 & 0 \\ 
\displaystyle   \rho_L \, g \,  \int_{\Gamma_0}  {\rm d}\gamma  \,   
\widetilde{\alpha}  \,\smb & 0 & 0     \end{pmatrix} 
\monend 
and we obtain as previously an operator 
$ \,  Q_0 (\Omega, \, {\cal S}) \ni q \longmapsto 
K \, \smb \, q \in  Q_0 (\Omega, \, {\cal S}) .\,$ 
We introduce also a global  right hand side  vector  $ \, F(t) \, $:
%
%
\moneq \label{global-F} 
F(t) \, = \,   \begin{pmatrix}   0 \cr (m_S + m_L)  \, g_0  \,+\, R \cr
  m_L \, \ell_0 \times  g_0  \,\,+\,\,  (x_A-\xi)   \times   R     \end{pmatrix} \,  
\monend 
and the relation 
$\, F(t) \in  Q_0 (\Omega, \, {\cal S}) \,$ is natural. 
We remark   with these relatively complicated definitions
 (\ref {espace-Q0}),    
 (\ref{matrice-M}),  (\ref{matrice-K}),  (\ref{global-F}) that  
the global dynamical system composed by the relations 
(\ref{impulsion-solide-4}), (\ref{moment-cinetique-4}), 
(\ref{continuite-pression-4})
admits finally  a very simple form:  
\moneq \label{global-oscillateur} 
M \,\smb \,   {{{\rm d}^2 q }\over{{\rm d}t^2}}  \,\, + \,\, K  \,\smb \, q 
\,=\, F(t) \, . 
\monend 
This equation is the extension of the previous free fluid oscillators equation 
(\ref{sloshing-classique}) to the coupling with  the solid motion. 

%

\bigskip    \monitem   {    \bf    Properties of the    mass matrix  } 

\noindent

\noindent
The matrix $M$ defined  in (\ref{matrice-M})  is symmetric and ``positive definite''. 
We  have the following expression for the quadratic form: 
\moneqstar 
 \big( q  \,, \, M \, \smb \, q   \big) \,=\,  m_S \,   \mid \! \xi \! \mid ^2  
\,+\, ( \theta \,,\,  {\rm I}_S   \, \smb \, \theta ) 
\,+\, \rho_L \,  \int_{  \Omega }   \mid  \nabla \alpha \,\smb  \, \xi  \,+\,
 \nabla \beta \,\smb  \, \theta \,+\,  \nabla \psi    \mid ^2   
\,  {\rm d} \gamma \, . 
\monendstar 
In other words,   we have the expression 
%
%
$ \,\,  T \,= \, {1 \over 2} \, \big(\,   {{{\rm d} q }\over{{\rm d}t}} \,, \, M \, \smb \, 
   {{{\rm d} q }\over{{\rm d}t}} \, \big)  \, \, $ 
%
for  the   kinetic energy developed in (\ref{kinetic-1011}). 
%
The proof of this proposition is detailed in \cite{DS14}. 


%
\bigskip   \noindent $\bullet$ \quad 
%
We consider  now the same questions for the rigidity operator $\, K $. 
We recall that the tangential coordinates $\, X_j \, $  
on the linearized free surface $ \, \Gamma_0 \,$ 
such that  $ \, x =   \sum_{j=1}^{3} X_j \, \varepsilon_j  \, $
satisfy  
 $ \, \int_{  \Gamma_0 }   X_j   \, {\rm d} \gamma  =  0 \,\,  $ for 
$ j = 1, \, 2 \, . $ 
We introduce a length $ \, a \, $ characteristic of this surface $ \, \Gamma_0 .\,$ 
Precisely, we suppose that  
\moneq \label{taille-caract}   
\int_{  \Gamma_0 }    \mid \!   X_j  \! \mid ^2   \, {\rm d} \gamma   \,\leq \, a^4  
\,, \quad j=1, \, 2 \, . 
\monend  
We introduce also the $ \, {\rm L}^2 \, $ norm $ \,  \displaystyle  \, \, 
 \mid \! \mid \! \eta  \! \mid   \! \mid \, \equiv $ 
$ \, \sqrt{ \int_{  \Gamma_0 }    \mid \!  \eta  \! \mid ^2   \, {\rm d} \gamma } \, $  
of the free surface, 
%
%
in coherence with the scalar product proposed in the relation (\ref{produit-scalaire}).

%
\bigskip   \noindent $\bullet$ \quad    {    \bf Proposition 8. 
 Properties of the rigidity   matrix    } 
    
\noindent     
The matrix $K$ is symmetric:  
%
$\,\,  \big( q  \,, \, K \, \smb \, q'   \big) \,= $ 
$  \,   \big(  K \, \smb \, q  \,, \, q'   \big) \, \,$ 
%
for arbitrary global vectors  $ \, q \, $ and  $ \, q' \, $
in the space  $ \, Q_0 (\Omega, \, {\cal S}) . \,$
The matrix $K$ is   positive:  
$ \,\,  \big( q  \,, \, K \, \smb \, q   \big) \,\geq \, 0  \,\,   $   
 if the  rotation $ \, \theta \, $ of the solid is sufficiently small 
relatively to the mean quadratic value of the free surface, {\it id est} 
\moneq \label{angle-petit}   
 \mid \! \theta_1  \! \mid  + \mid \! \theta_2  \! \mid 
\,\, \leq \, {{1}\over{2 \, a^2}} \, 
 \mid \! \mid \! \eta  \! \mid   \! \mid   \, . 
\monend 
This relation is quite precise concerning the validity of linearity hypotheses. 
%

\bigskip   \noindent $\bullet$ \quad    {    \bf  Proof of Proposition 8. }  
 
\noindent   
The symmetry of the matrix $K$ is elementary to establish.  We 
%
refer to \cite{DS14}. 
We have also: 

\smallskip  \noindent  $ \,  \displaystyle   
\big( q \,, \, K \,\smb\, q \big)   \,= \, 
  \rho_L \, g \, \Big(  2 \, \int_{  \Gamma_0 }  \! \theta \smb  \,  \widetilde{\alpha}   \,
 \eta \, \, {\rm d}   \gamma 
  \,+\,    \mid \! \mid \! \eta  \! \mid   \! \mid ^2  \, \Big) = 
   \rho_L \, g \,\,  \Big[  \, 2  \, \int_{  \Gamma_0 } \! \eta \, \big( X_2 \, \theta_1 - 
    X_1 \, \theta_2 \big)  \, {\rm d}   \gamma 
  \,+\,    \mid \! \mid \! \eta  \! \mid   \! \mid ^2  \, \Big] \, . $

 \smallskip   \noindent  Then 

 \smallskip  \noindent  $ \,  \displaystyle    
  \vert   \int_{  \Gamma_0 } \eta \,  X_2 \, \theta_1 
 \, {\rm d}   \gamma   \,    \vert   \,\, \leq \,\,   
\,\,\,  \mid\!  \theta_1 \!\mid 
\int_{  \Gamma_0 }  \mid\!  \eta  \!\mid  \,\,   \mid\!    X_2  \!\mid 
\, {\rm d}   \gamma  \, $ 
$\hfill$ because $ \,  \theta_1 \,$ is a constant on $ \, \Gamma_0  $  
 
 \smallskip   \qquad    $ \,\,\,   \displaystyle \, \leq \, 
\,\,\, \mid \!  \theta_1 \! \mid  
\,\,   \mid \! \mid \! \eta  \! \mid   \! \mid 
\,\,   \mid \! \mid \! X_2  \! \mid   \! \mid   $ 
  $ \hfill $ using the   Cauchy-Schwarz inequality  

 \smallskip   \qquad    $ \,\,\,   \displaystyle \, \leq \,  
\,\,\, \mid \!  \theta_1 \! \mid  
\,\,   \mid \! \mid \! \eta  \! \mid   \! \mid \,\,  a^2 $ 
   $ \hfill $ by hypothesis (\ref{taille-caract})

 \smallskip  \noindent   and the analogous inequality 
 \quad   $ \,  \displaystyle \,   \mid  \int_{  \Gamma_0 } \eta \,  X_1 \, \theta_2 
 \, {\rm d}   \gamma   \mid   \,\, \leq \,\, 
\,\,\, \mid \!  \theta_2 \! \mid  
\,\,   \mid \! \mid \! \eta  \! \mid   \! \mid \,\,  a^2 $ \quad 
for the other component. 
We deduce from the previous assessment  the minoration:

 \smallskip  \noindent  $ \,  \displaystyle    
\big( q \,, \, K \,\smb\, q \big) \, \, \geq \, \,  \rho_L \, g \, \Big[ 
\,      \mid \! \mid \! \eta  \! \mid   \! \mid ^2   \,\,-\,\, 
2 \, a^2 \,   \mid \! \mid \! \eta  \! \mid   \! \mid \,\ \big( 
 \mid \!  \theta_1 \! \mid  +  \mid \!  \theta_2 \! \mid  \big) \, \Big]  $

 \smallskip   \qquad \quad     $    \displaystyle \,\,\, \geq \, \,  \rho_L \, g \,  
\mid \! \mid \! \eta  \! \mid   \! \mid  \,\, \Big[ 
\,      \mid \! \mid \! \eta  \! \mid   \! \mid    \,\,-\,\, 
2 \, a^2   \,\ \big( 
 \mid \!  \theta_1 \! \mid  +  \mid \!  \theta_2 \! \mid  \big) \, \Big] $

 \smallskip  \noindent    and this expression is positive when 
 \quad      $ \,  \displaystyle  
 \mid \! \theta_1  \! \mid  + \mid \! \theta_2  \! \mid 
\,\, \leq \, {{1}\over{2 \, a^2}} \, 
 \mid \! \mid \! \eta  \! \mid   \! \mid   \, \,  $ \quad
which is exactly the hypothesis (\ref{angle-petit}). 
   The proposition is established. 
$ \hfill $  $\square$  


\bigskip   \newpage  \monitem   {    \bf    Lagrangian function for the coupled system  } 

\noindent 
With the reduction of the coupled sloshing problem to the unknown $ \, q \equiv (  \eta ,\,  \xi, 
\theta  ) \in   Q_0 (\Omega, \, {\cal S}) \,$ 
we first specify  the energies according to this global field. 
The conservation of energy (\ref{conservation-energie}) 
has been established again from the compact form (\ref{global-oscillateur})
of the evolution equation. 
%
If the  external  force $ \, R(t) \,$ is equal to zero, it is natural
to introduce the Lagrangian $ \, {\cal L} \, $ according to the 
usual definition:
\moneq \label{lagrangien}  
 {\cal L} \,=\, T - (U + V) \, . 
  \monend 
Then this Lagrangian is a functional of the state $ \, q \,$ defined in 
(\ref{global-vector}) and of its first time derivative. 
We have the final Proposition:

\bigskip   \noindent $\bullet$ \quad    {    \bf Proposition 9. 
 Euler-lagrange equations  } 

\noindent     
With the above notations when the right hand side   $\, F(t) \,$
is reduced to the gravity term,  the equations of motion (\ref{global-oscillateur}) 
take the form 
\moneqstar 
{{{\rm d}}\over{{\rm d}t}} \, 
\Big(  {{\partial  {\cal L}}\over {\partial 
\big(  {{{\rm d} q}\over{{\rm d} t}} \big) }} \Big)  \,=\, 
 {{\partial  {\cal L}}\over {\partial q}}  \, .  
  \monendstar  
The proof of this proposition is elementary. 
We omit it and refer to  \cite{DS14}. 

\monitem 
With this general framework, the Lagrangian formulation is 
simple to use. 
It is sufficient for the applications 
to evaluate carefully the Lagrangian $ \,  {\cal L} \,$
given by the relations  (\ref{kinetic-1011}),  
(\ref{poten-1011}), (\ref{exte-1011}) and  (\ref{lagrangien}). 


\bigskip \bigskip   \noindent {\bf \large   Conclusion   }
 
\noindent 
In this contribution, we started from our industrial practice of 
sloshing for rigid bodies submitted to an acceleration.
We first set the importance of the irrotational hypothesis
of the flow in the external Galilean reference frame. 
Then we derived carefully the mechanics  
of the solid motion (conservation of momentum and conservation of kinetic momentum)
and of the fluid motion (Laplace equation for the velocity potential), 
with a particular emphasis for  the coupling with the continuity
of the normal velocity field and the continuity of pressure across the 
fluid surface. 
A first difficulty is   the representation of the solid rotational 
  velocity vector field with potential functions. 
This can be achieved with the Stokes-Zhukovsky vector fields that are 
particular harmonic functions associated to the geometry of the fluid.
Efficient numerical methods like integral methods
 (see {\it e.g.} \cite{Ne01}) could be used to go one step further. 
A much well  known mathematical difficulty is the reduction of the 
fluid problem to a Neumann to Dirichlet operator for
the Laplace equation. 
The use of integral methods is also natural for this kind of coupling
(see {\it e.g.} \cite{MR65} and \cite {MO92}).  
%
In particular, the integral algorithm  used {\it e.g.} in 
our contribution  \cite{BCD3T15} is appropriate for such numerical computation.
 The  degrees of freedom of both Stokes-Zhukovsky and modal functions 
are located on the interface between solid and liquid and the total
computational cost of such approach is reasonable. 
%
Last but not least, we have derived a general expression for the Lagrangian of this
coupled system. 
The next step is to look to simplified systems and confront our rigorous mathematical
analysis with the state of the art in the engineering community. 
In particular, we are interested in developing  
appropriate methodologies to define  equivalent 
simplified mechanical systems as the ones presented in~\cite {NA66}. 
We plan also to apply our formulation with a Neumann to Dirichlet operator 
with boundary element methods.

\bigskip \bigskip  \noindent {\bf \large Acknowledgments}   

\noindent  
The authors  thank their colleagues of   Airbus Defence and Space 
 Christian Le~Noac'h   for  enthusiastic interaction, 
Gerald Pigni\'e for helpful comments all along this work, and 
Fran\c cois Coron for suggesting us to study this problem.  
 They thank also Roger Ohayon
 of   Conservatoire National des Arts et M\'etiers in Paris for a detailed 
 bibliography transmitted  in January 2008. 
A special thanks to Antoine Mareschal of Institut Polytechnique des Sciences 
Avanc\'ees for his internship in Les Mureaux in 2013. 
Last but not least, 
the authors thank the referees for  very constructive remarks. 
Some of them  have been   
incorporated into  the present edition of the article.


\medskip

\end{document}